\documentclass[11pt,journal,onecolumn,draftcls]{IEEEtran}
\IEEEoverridecommandlockouts
\usepackage[table]{xcolor}
\usepackage{array}
\usepackage{bbding}
\usepackage{cite}
\usepackage{graphicx}
\usepackage{amssymb}
\usepackage[cmex10]{amsmath}
\usepackage{amsfonts}
\usepackage{algorithmic}
\usepackage{algorithm}
\usepackage{mdwmath}
\usepackage{mdwtab}
\usepackage{eqparbox}
\usepackage[caption=false,font=footnotesize]{subfig}
\usepackage{fixltx2e}
\usepackage{stfloats}
\usepackage{dsfont}
\usepackage{mathrsfs}
\usepackage{bbm}
\usepackage{bm}
\usepackage{threeparttable}
\usepackage{multirow}

\newtheorem{assumption}{Assumption}

\newcommand{\qed}{\nobreak \ifvmode \relax \else
      \ifdim\lastskip<1.5em \hskip-\lastskip
      \hskip1.5em plus0em minus0.5em \fi \nobreak
      \vrule height0.75em width0.5em depth0.25em\fi}

\def\mc{\mathcal}
\def\bs{\boldsymbol}
\def\defeq{\triangleq}
\def\st{{\mathrm{subject\;to}}}
\DeclareMathOperator*{\minimize}{\mathrm{minimize}}

\def\ds{\mathds}

\def\wt{\widetilde}
\def\wh{\widehat}
\def\ol{\overline}
\def\E{{\mathbb{E}}}
\def\col{{\mathrm{col}}}
\def\diag{{\mathrm{diag}}}

\def\Tr{{\mathrm{Tr}}}
\def\vecm{{\mathrm{vec}}}

\def\T{{\mathsf{T}}}

\def\N{{\mathcal{N}}}
\def\MSD{{\textrm{MSD}}}
\def\EMSE{{\textrm{EMSE}}}
\def\atc{{\textrm{atc}}}
\def\cta{{\textrm{cta}}}
\def\opt{{\textrm{opt}}}
\def\unf{{\textrm{unf}}}
\def\blk{{\textrm{blk}}}
\def\inc{{\textrm{inc}}}
\def\ind{{\textrm{ind}}}

\def\lambdamin{\lambda_{\mathrm{min}}}
\def\smallvskip{\vskip 0.5em}

\begin{document}

\title{Performance Limits for Distributed Estimation Over LMS Adaptive Networks}

\author{Xiaochuan~Zhao,~\IEEEmembership{Student~Member,~IEEE,}
        and Ali~H.~Sayed,~\IEEEmembership{Fellow,~IEEE}
\thanks{This work was supported in part by NSF grants CCF-0942936 and CCF-1011918. A short early version of this work was presented in IEEE International Conference on Acoustics, Speech, and Signal Processing (ICASSP), Prague, Czech Republic, May 2011 \cite{Zhao11ICASSP}.}
\thanks{The authors are with Department of Electrical Engineering, University of California, Los Angeles, CA 90095
Email: \{xzhao,~sayed\}@ee.ucla.edu.}
}


\maketitle

\begin{abstract}
In this work we analyze the mean-square performance of different strategies for distributed estimation over least-mean-squares (LMS) adaptive networks. The results highlight some useful properties for distributed adaptation in comparison to fusion-based centralized solutions. The analysis establishes that, by optimizing over the combination weights, diffusion strategies can deliver lower excess-mean-square-error than centralized solutions employing traditional block or incremental LMS strategies. We first study in some detail the situation involving combinations of two adaptive agents and then extend the results to generic $N$-node ad-hoc networks. In the later case, we establish that, for sufficiently small step-sizes, diffusion strategies can outperform centralized block or incremental LMS strategies by optimizing over left-stochastic combination weighting matrices. The results suggest more efficient ways for organizing and processing data at fusion centers, and present useful adaptive strategies that are able to enhance performance when implemented in a distributed manner.
\end{abstract}



\newpage 

\section{Introduction}
\IEEEPARstart{T}{his} work examines the dynamics that results when adaptive nodes are allowed to interact with each other. Through cooperation, some interesting behavior occurs that is not observed when the nodes operate independently. For example, if one adaptive agent has worse performance than another independent adaptive agent, can both agents cooperate with each other in such a manner that the performance of \emph{both} agents improves? What if $N$ agents are interacting with each other? Can all agents improve their performance relative to the non-cooperative case even when some of them are noisier than others? Does cooperation need to be performed in a centralized manner or is distributed cooperation sufficient to achieve this goal? Starting with two adaptive nodes, we derive analytical expressions for the mean-square performance of the nodes under some conditions on the measurement data. The expressions are then used to compare the performance of various (centralized and distributed) adaptive strategies. The analysis reveals a useful fact that arises as a result of the cooperation between the nodes; it establishes that, by optimizing over the combination weights, diffusion least-mean-squares (LMS) strategies for distributed estimation can deliver lower excess-mean-square-error (EMSE) than a centralized solution employing traditional block or incremental LMS strategies. We first study in some detail the situation involving combinations of two adaptive nodes for which the performance levels can be characterized analytically. Subsequently, we extend the conclusion to $N$-node ad-hoc networks. Reference \cite{Sayed13Chapter} provides an overview of diffusion strategies for adaptation and learning over networks.

It is worth noting that the performance of diffusion algorithms was already studied in some detail in the earlier works \cite{Lopes08TSP,Cattivelli10TSP}. These works derived expressions for the network EMSE and mean-square-deviation (MSD) in terms of the combination weights that are used during the adaptation process. The results in \cite{Lopes08TSP,Cattivelli10TSP} were mainly concerned in comparing the performance of diffusion (i.e., distributed cooperative) strategies with \emph{non-cooperative} strategies. In the cooperative case, nodes share information with each other, whereas they behave independently of each other in the non-cooperative case. In the current work, we are instead interested in comparing diffusion or \emph{distributed} cooperative strategies against \emph{centralized} (as opposed to non-cooperative) strategies. In the centralized framework, a fusion center has access to all data collected from across the network, whereas in the non-cooperative setting nodes have access only to their individual data. Therefore, finding conditions under which diffusion strategies can perform well in comparison to centralized solutions is generally a demanding task.

We start our study by considering initially the case of two interacting adaptive agents. Though structurally simple, two-node networks are important in their own right. For instance, two-antenna receivers are prevalent in many communication systems. The data received by the antennas could either be transferred to a central processor for handling or processed cooperatively and locally at the antennas. Which mode of processing can lead to better performance and how? Some of the results in this article help provide answers to these questions. In addition, two-node adaptive agents can serve as good models for how estimates can be combined at master nodes that connect larger sub-networks together. There has also been useful work in the literature on examining the performance of combinations of two adaptive filters \cite{Arenas05TIM,Arenas06TSP,Mandic07ICASSP,Silva08TSP,Candido10TSP}. The main difference between two-node adaptive networks and combinations of two adaptive filters is that in the network case the measurement and regression data are fully distributed and also different across nodes, whereas the filters share the same measurement and regression data in filter combinations \cite{Arenas05TIM,Arenas06TSP,Mandic07ICASSP,Silva08TSP,Candido10TSP}. For this reason, the study of adaptive networks is more challenging and their dynamics is richer.

The results in this work will reveal that distributed diffusion LMS strategies can outperform centralized block or incremental LMS strategies through proper selection of the combination weights. The expressions for the combination weights end up depending on knowledge of the noise variances, which are generally unavailable to the nodes. Nevertheless, the expressions suggest a useful adaptive construction. Motivated by the analysis, we propose an adaptive method for adjusting the combination weights by relying solely on the available data. Simulation results illustrate the findings.

\emph{Notation}: We use lowercase letters to denote vectors, uppercase letters for matrices, plain letters for deterministic variables, and boldface letters for random variables. We also use $(\cdot)^{\T}$ to denote transposition, $(\cdot)^*$ for conjugate transposition, $(\cdot)^{-1}$ for matrix inversion, $\Tr(\cdot)$ for the trace of a matrix, $\rho(\cdot)$ for the spectral radius of a matrix, $\otimes$ for the Kronecker product, ${\mathrm{vec}}(A)$ for stacking the columns of $A$ on top of each other, and $\diag(A)$ for constructing a vector by using the diagonal entries of $A$. All vectors in our treatment are column vectors, with the exception of the regression vectors, ${\bs{u}}_{k,i}$, which are taken to be row vectors for convenience of presentation.

\subsection{Non-Cooperative Adaptation by Two Nodes}
We refer to the two nodes as nodes 1 and 2. Both nodes are assumed to measure data that satisfy a linear regression model of the form:
\begin{align}
\label{eqn:linearmodel}
  {\bs{d}}_{k}(i)={\bs{u}}_{k,i}w^o+{\bs{v}}_{k}(i)
\end{align}
for $k=1,2$, where $w^o$ is a deterministic but unknown $M\times1$ vector, ${\bs{d}}_k(i)$ is a random measurement datum at time $i$, ${\bs{u}}_{k,i}$ is a random $1\times M$ regression vector at time $i$, and ${\bs{v}}_{k}(i)$ is a random noise signal also at time $i$. We adopt the following assumptions on the statistical properties of the data $\{{\bs{u}}_{k,i},{\bs{v}}_{k}(i)\}$.

\begin{assumption}[Statistical properties of the data]
\label{asm:all} \hskip 0.5em  \;\;
\begin{enumerate}
  \item The regression data ${\bs{u}}_{k,i}$ are temporally white and spatially independent random variables with zero mean and uniform covariance matrix $R_{u,k}\defeq\E{\bm{u}}_{k,i}^*{\bm{u}}_{k,i}>0$.
  \item The noise signals ${\bs{v}}_{k}(i)$ are temporally white and spatially independent random variables with zero mean and variances $\sigma_{v,k}^2$.
  \item The regressors ${\bs{u}}_{k,i}$ and noise signals ${\bs{v}}_{l}(j)$ are mutually-independent for all $k$ and $l$, $i$ and $j$. \hfill \IEEEQED
\end{enumerate}
\end{assumption}

\noindent It is worth noting that we do not assume Gaussian distributions for either the regressors or the noise signals. We note that the temporal independence assumption on the regressors may be invalid in general, especially for tapped-delay implementations where the regressions at each node would exhibit a shift structure. However, there have been extensive studies in the stochastic approximation literature showing that, for stand-alone adaptive filters, results based on the temporal independence assumption, such as \eqref{eqn:EMSElms} and \eqref{eqn:MSDlms} further ahead, still match well with actual filter performance when the step-size is sufficiently small \cite{Ljung77TAC1,Ljung77TAC2,Benveniste90,MacChi95,Solo95,Kushner03,Sayed08}. Thus, we shall adopt the following assumption throughout this work.

\begin{assumption}[Small step-sizes]
\label{asm:smallstepsize}
The step-sizes are sufficiently small, i.e., $\mu_k\ll1$, so that terms depending on higher-order powers of the step-sizes can be ignored, and such that the adaptive strategies discussed in this work are mean-square stable (in the manner defined further ahead). \hfill \IEEEQED
\end{assumption}

\begin{figure}[t]
  \centering
  \includegraphics[width=2.8in]{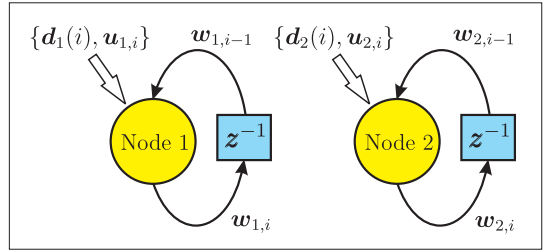}
  \caption{Nodes 1 and 2 process the data independently by means of two local LMS filters.}
  \label{fig:indLMS}
\end{figure}

We are interested in the situation where one node is less noisy than the other. Thus, without loss of generality, we assume that the noise variance of node 2 is less than that of node 1, i.e.,
\begin{equation}
\boxed{
\sigma_{v,2}^2<\sigma_{v,1}^2
}\end{equation}
The nodes are interested in estimating the unknown parameter $w^o$. Assume initially that each node $k$ independently adopts the famed LMS algorithm \cite{Widrow85,Haykin02,Sayed08} to update its weight estimate (as illustrated in Fig. \ref{fig:indLMS}) according to the following rule:
\begin{align}
\label{eqn:standaloneLMS}
{\bs{w}}_{k,i}={\bs{w}}_{k,i-1}+\mu_k{\bs{u}}_{k,i}^*\left[{\bs{d}}_{k}(i)-{\bs{u}}_{k,i}{\bs{w}}_{k,i-1}\right]
\end{align}
for $k=1,2$, where $\mu_k$ is a positive constant step-size parameter. The steady-state performance of an adaptive algorithm is usually assessed in terms of its mean-square error (MSE), EMSE, and MSD, which are defined as follows. If we introduce the error quantities:
\begin{align}
\label{eqn:errorsdef1}
{\bs{e}}_{k}(i)&\defeq{\bs{d}}_{k}(i)-{\bs{u}}_{k,i}{\bs{w}}_{k,i-1}\\
\label{eqn:errorsdef2}
{\widetilde{\bs{w}}}_{k,i}&\defeq w^o-{\bs{w}}_{k,i}\\
\label{eqn:errorsdef3}
{\bs{e}}_{a,k}(i)&\defeq{\bs{u}}_{k,i}{\widetilde{\bs{w}}}_{k,i-1}
\end{align}
then the MSE, EMSE, and MSD for node $k$ are defined as the following steady-state values:
\begin{align}
{\textrm{MSE}}_k&\defeq\lim_{i\rightarrow\infty}\E|{\bs{e}}_{k}(i)|^2\\
{\EMSE}_k&\defeq\lim_{i\rightarrow\infty}\E|{\bs{e}}_{a,k}(i)|^2\\
{\MSD}_k&\defeq\lim_{i\rightarrow\infty}\E\|{\widetilde{\bs{w}}}_{k,i}\|^2
\end{align}
where the notation $\|\cdot\|$ denotes the Euclidean norm of its vector argument. Substituting expression \eqref{eqn:linearmodel} into the definition for ${\bs{e}}_{k}(i)$ in \eqref{eqn:errorsdef1}, it is easy to verify that the errors $\{{\bs{e}}_{k}(i),{\bs{e}}_{a,k}(i)\}$ are related as follows:
\begin{align}
{\bs{e}}_{k}(i)={\bs{e}}_{a,k}(i)+{\bm{v}}_{k}(i)
\end{align}
for $k=1,2$. Since the terms ${\bm{v}}_{k}(i)$ and ${\bs{e}}_{a,k}(i)$ are independent of each other, it readily follows that the MSE and EMSE performance measures at each node are related to each other through the noise variance:
\begin{align}
{\textrm{MSE}}_k={\EMSE}_k+\sigma_{v,k}^2
\end{align}
Therefore, it is sufficient to examine the EMSE and MSD as performance metrics for adaptive algorithms. Under Assumption \ref{asm:smallstepsize}, the EMSE and MSD of each LMS filter in \eqref{eqn:standaloneLMS} are known to be well approximated by \cite{Widrow76Proc,Horowitz81TSP,Jones82TIT,Gardner84SP,Widrow85,Feuer85TSP,Foley88TSP,Haykin02,Sayed08}:
\begin{equation}
\label{eqn:EMSElms}
\boxed{
{\EMSE}_k\approx\frac{1}{2}\,\mu_k\sigma_{v,k}^2\Tr(R_{u,k})
}\end{equation}
and
\begin{equation}
\label{eqn:MSDlms}
\boxed{
{\MSD}_k\approx\frac{1}{2}\,\mu_k\sigma_{v,k}^2M
}\end{equation}
for $k=1,2$. To proceed, we further assume that both nodes employ the same step-size and observe data arising from the same underlying distribution.

\begin{assumption}[Uniform step-sizes and data covariance]
\label{asm:uniform}
It is assumed that both nodes employ identical step-sizes, i.e., $\mu_1=\mu_2=\mu$, and that they observe data arising from the same statistical distribution, i.e., $R_{u,1}=R_{u,2}=R_u$.  \hfill \IEEEQED
\end{assumption}

\noindent Under Assumption \ref{asm:uniform}, expression \eqref{eqn:EMSElms} confirms the expected result that node 2 will achieve a lower EMSE than node 1 because node 2 has lower noise variance than node 1. The interesting question that we would like to consider is whether it is possible to improve the EMSE performance for \emph{both} nodes if they are allowed to cooperate with each other in some manner. The arguments in this work will answer this question in the affirmative and will present \emph{distributed} cooperative schemes that are able to achieve this goal, not only for two-node networks but also for $N$-node ad-hoc networks (see Sec. VI).

\section{Two Centralized Adaptive Algorithms}
\begin{figure}[t]
\centerline{
\subfloat[Block LMS adaptation.]
{\includegraphics[width=2in]{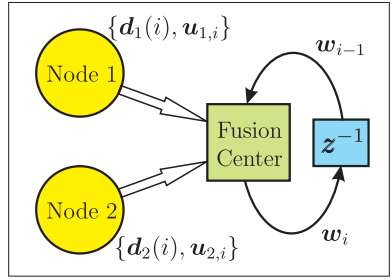}
\label{fig:vecLMS}}}
\centerline{
\subfloat[Incremental LMS adaptation.]
{\includegraphics[width=3in]{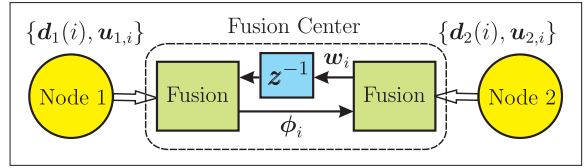}
\label{fig:incLMS}}}
\caption{Two centralized strategies using data from nodes at a fusion center.}
\label{fig:centralized}
\end{figure}

One form of cooperation can be realized by connecting the two nodes to a fusion center, which would collect the data from the nodes and and use them to estimate $w^o$. Fusion centers are generally more powerful than the individual nodes and can, in principle, implement more sophisticated estimation procedures than the individual nodes. In order to allow for a fair comparison between implementations of similar nature at the fusion center and remotely at the nodes, we assume that the fusion center is limited to implementing LMS-type solutions as well, albeit in a centralized manner. In this work, the fusion center is assumed to operate on the data in one of two ways. The first method is illustrated in Fig. \ref{fig:vecLMS} and we refer to it as \emph{block LMS}. In this method, the fusion center receives data from the nodes and updates its estimate for $w^o$ according to the following:
\begin{align}
\label{eqn:blockLMS}
{\bs{w}}_{i}={\bs{w}}_{i-1}+\mu'\begin{bmatrix}
{\bs{u}}_{1,i} \\
{\bs{u}}_{2,i} \\
\end{bmatrix}^*\left(\begin{bmatrix}
{\bs{d}}_{1}(i) \\
{\bs{d}}_{2}(i) \\
\end{bmatrix}-\begin{bmatrix}
{\bs{u}}_{1,i} \\
{\bs{u}}_{2,i} \\
\end{bmatrix}{\bs{w}}_{i-1}\right)
\end{align}
with a constant positive step-size $\mu'$. The second method is illustrated in Fig. \ref{fig:incLMS} and we refer to it as \emph{incremental LMS}. In this method, the fusion center still receives data from the nodes but operates on them sequentially by incorporating one set of measurements at a time as follows:
\begin{equation}
\label{eqn:incrementalLMS}
\left\{\begin{aligned}
{\bs{\phi}}_{i}&={\bs{w}}_{i-1}+\mu'{\bs{u}}_{1,i}^*\left[{\bs{d}}_{1}(i)-{\bs{u}}_{1,i}{\bs{w}}_{i-1}\right]\\
{\bs{w}}_{i}&={\bs{\phi}}_{i}+\mu'{\bs{u}}_{2,i}^*\left[{\bs{d}}_{2}(i)-{\bs{u}}_{2,i}{\bs{\phi}}_{i}\right]\\
\end{aligned}\right.
\end{equation}
We see from \eqref{eqn:incrementalLMS} that the fusion center in this case first uses the data from node 1 to update ${\bs{w}}_{i-1}$ to an intermediate value ${\bs{\phi}}_{i}$, and then uses the data from node 2 to get ${\bs{w}}_{i}$. Method \eqref{eqn:incrementalLMS} is a special case of the incremental LMS strategy introduced and studied in \cite{Lopes07TSP} and is motivated by useful incremental approaches to distributed optimization \cite{Polyak73ARC,Bertsekas97JOP,Bertsekas97,Nedic00,Nedic01JOP,Rabbat05JSAC}. We observe from \eqref{eqn:blockLMS} and \eqref{eqn:incrementalLMS} that in going from ${\bs{w}}_{i-1}$ to ${\bs{w}}_i$, the block and incremental LMS algorithms employ two sets of data for each such update; in comparison, the conventional LMS algorithm used by the stand-alone nodes in \eqref{eqn:standaloneLMS} employs one set of data for each update of their respective weight estimates.

We define the EMSE and MSD for block LMS \eqref{eqn:blockLMS} and incremental LMS \eqref{eqn:incrementalLMS} as follows:
\begin{align}
\label{eqn:EMSEblkdef}
{\EMSE}_{\mathrm{blk/inc}}&\defeq\frac{1}{2}\lim_{i\rightarrow\infty}\E\|{\bs{e}}_{a,i}\|^2\\
\label{eqn:MSDblkdef}
{\MSD}_{\mathrm{blk/inc}}&\defeq\lim_{i\rightarrow\infty}\E\|{\widetilde{\bs{w}}}_{i}\|^2
\end{align}
where the a priori error ${\bs{e}}_{a,i}$ is now a $2\times1$ vector:
\begin{align}
\label{eqn:erroraidefinc}
{\bs{e}}_{a,i}\defeq\begin{bmatrix}
{\bs{u}}_{1,i}{\wt{\bm{w}}}_{i-1} \\
{\bs{u}}_{2,i}{\wt{\bm{w}}}_{i-1}
\end{bmatrix}
\end{align}
Note that in \eqref{eqn:EMSEblkdef} we are scaling the definition of the EMSE by $1/2$ because the squared Euclidean-norm in \eqref{eqn:EMSEblkdef} involves the sum of the two error components from \eqref{eqn:erroraidefinc}. We shall explain later in Sec. VI that in order to ensure a fair comparison of the performance of the various algorithms (including non-cooperative, distributed, and centralized), we will need to set the step-size as (see \eqref{eqn:mu1andmu2})
\begin{align}
\label{eqn:stepsize}
\mu'=\frac{\mu}{2}
\end{align}
This normalization will help ensure that the rates of convergence of the various strategies that are being compared are similar.

Now, compared to the non-cooperative method \eqref{eqn:standaloneLMS} where the nodes act individually, it can be shown that the two centralized algorithms \eqref{eqn:blockLMS} and \eqref{eqn:incrementalLMS} lead to improved mean-square performance (the arguments further ahead in Sec. IV-F establish this conclusion among several other properties). Specifically, the EMSE obtained by the two centralized algorithms \eqref{eqn:blockLMS} and \eqref{eqn:incrementalLMS} will be smaller than the average EMSE obtained by the two non-cooperative nodes in \eqref{eqn:standaloneLMS}. The question that we would like to explore is whether distributed cooperation between the nodes can lead to superior performance even in comparison to the centralized algorithms \eqref{eqn:blockLMS} and \eqref{eqn:incrementalLMS}. To address this question, we shall consider distributed LMS algorithms of the diffusion-type from \cite{Lopes08TSP,Cattivelli10TSP}, and which are further studied in \cite{Li09SSP,Cattivelli10TAC,Takahashi10TSP,Takahashi10ICASSP,Chouvardas11TSP,Tu11CAMSAP,Tu12SSP,Tu12TSP,Zhao12TSP}. Reference \cite{Sayed13Chapter} provides an overview of diffusion strategies. Adaptive diffusion strategies have several useful properties: they are scalable and robust, enhance stability, and enable nodes to adapt and learn through localized interactions. There are of course other useful algorithms for distributed estimation that rely instead on consensus-type strategies, e.g., \cite{Schizas09TSP,Dimakis10PROC,Kar11TSP}. Nevertheless, diffusion strategies have been shown to lead to superior mean-square-error performance in comparison to consensus-type strategies (see, e.g., \cite{Tu12SSP,Tu12TSP}). For this reason, we focus on comparing adaptive diffusion strategies with the centralized block and incremental LMS approaches. The arguments further ahead will show that diffusion algorithms are able to exploit the spatial diversity in the data more fully than the centralized implementations and can lead to better steady-state mean-square performance than the block and incremental algorithms \eqref{eqn:blockLMS} and \eqref{eqn:incrementalLMS}, when all algorithms converge in the mean-square sense at the same rate. We shall establish these results initially for the case of two interacting adaptive agents, and then discuss the generalization for $N$-node networks in Sec. VI.

\section{Adaptive Diffusion Strategies}
\begin{figure}[t]
\centerline{
\subfloat[CTA diffusion adaptation.]
{\includegraphics[width=3in]{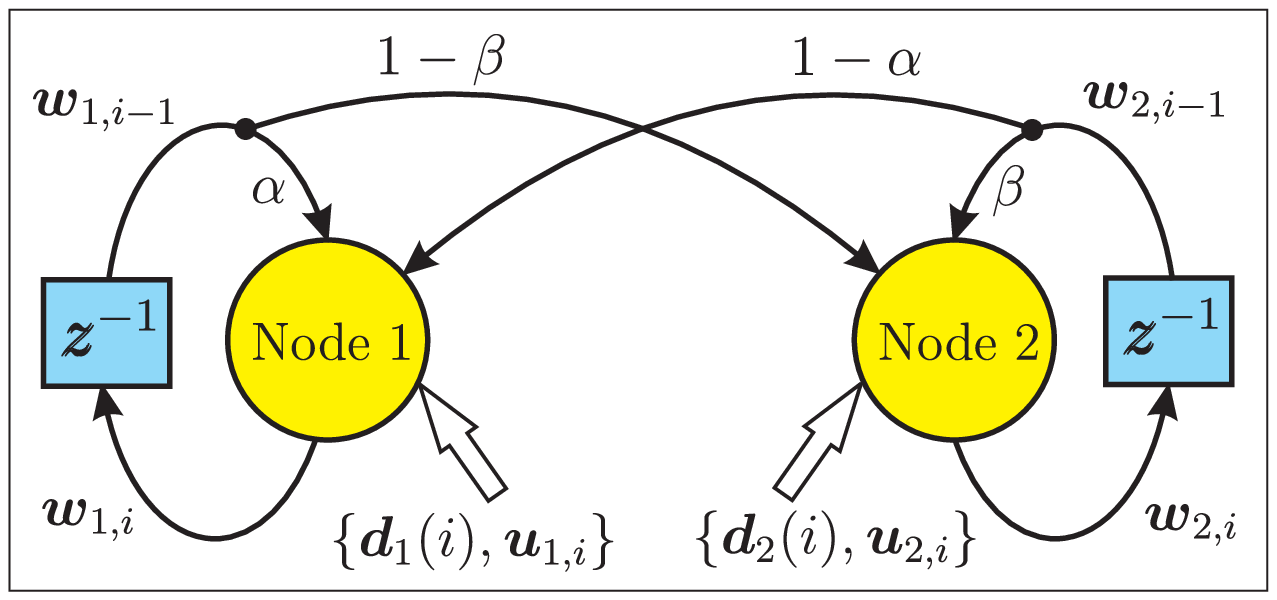}
\label{fig:diffLMScta}}}
\centerline{
\subfloat[ATC diffusion adaptation.]
{\includegraphics[width=3in]{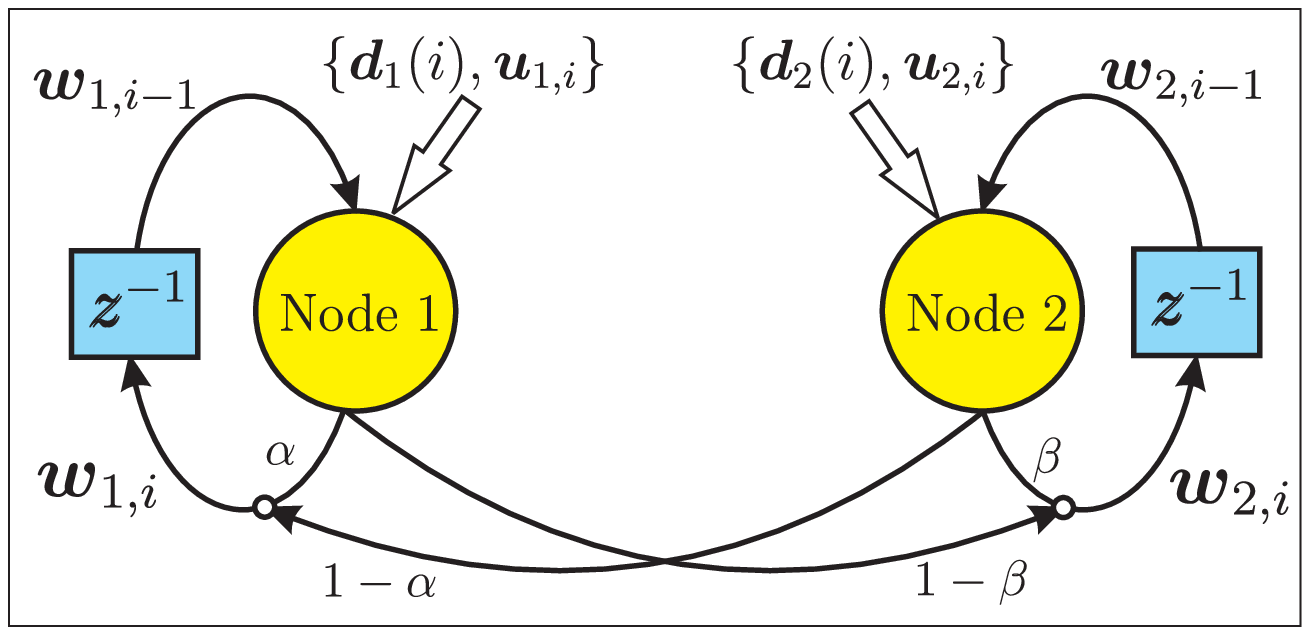}
\label{fig:diffLMSatc}}}
\caption{Two diffusion strategies using combination coefficients $\{\alpha,1-\alpha,\beta,1-\beta\}$.}
\label{fig:diffusion}
\end{figure}

Diffusion LMS algorithms are distributed strategies that consist of two steps \cite{Sayed13Chapter,Lopes08TSP,Cattivelli10TSP}: updating the weight estimate using local measurement data (the adaptation step) and aggregating the information from the neighbors (the combination step). According to the order of these two steps, diffusion algorithms can be categorized into two classes: Combine-then-Adapt (CTA) (as illustrated in Fig. \ref{fig:diffLMScta}):
\begin{equation}
\label{eqn:CTAnodek}
\left\{\begin{aligned}
{\bs{\phi}}_{k,i-1}&=a_{1k}{\bs{w}}_{1,i-1}+a_{2k}{\bs{w}}_{2,i-1}\\
{\bs{w}}_{k,i}&={\bs{\phi}}_{k,i-1}+\mu_k{\bs{u}}_{k,i}^*\left[{\bs{d}}_{k}(i)-{\bs{u}}_{k,i}{\bs{\phi}}_{k,i-1}\right]
\end{aligned}\right.
\end{equation}
and Adapt-then-Combine (ATC) (as illustrated in Fig. \ref{fig:diffLMSatc}):
\begin{equation}
\label{eqn:ATCnodek}
\left\{\begin{aligned}
{\bs{\psi}}_{k,i}&={\bs{w}}_{k,i-1}+\mu_k{\bs{u}}_{k,i}^*\left[{\bs{d}}_{k}(i)-{\bs{u}}_{k,i}{\bs{w}}_{k,i-1}\right]\\
{\bs{w}}_{k,i}&=a_{1k}{\bs{\psi}}_{1,i}+a_{2k}{\bs{\psi}}_{2,i}
\end{aligned}\right.
\end{equation}
for $k=1,2$, where the $\{\mu_k\}$ are positive step-sizes and the $\{a_{lk}\}$ denote convex combination coefficients used by nodes 1 and 2. The coefficients are nonnegative and they satisfy
\begin{align}
a_{1k}\ge0,\qquad a_{2k}\ge0,\qquad a_{1k}+a_{2k}=1
\end{align}
for $k=1,2$. We collect these coefficients into a $2\times2$ matrix $A$ and denote them more compactly by $\{\alpha,1-\alpha\}$ for node 1 and $\{1-\beta,\beta\}$ for node 2:
\begin{align}
\label{eqn:A2def}
A\defeq\begin{bmatrix}
  a_{11} \;&\; a_{12} \\
  a_{21} \;&\; a_{22}
\end{bmatrix}=\begin{bmatrix}
  \alpha \;&\; 1-\beta \\
  1-\alpha \;&\; \beta
\end{bmatrix}
\end{align}
where $\alpha,\beta\in[0,1]$. Note that when $\alpha=\beta=1$, both CTA algorithm \eqref{eqn:CTAnodek} and the ATC algorithm \eqref{eqn:ATCnodek} reduce to the non-cooperative LMS update given by \eqref{eqn:standaloneLMS}; we shall exclude this case for diffusion algorithms. Observe that the order of adaptation and combination steps are different for CTA and ATC implementations. The ATC algorithm \eqref{eqn:ATCnodek} is known to outperform the CTA algorithm \eqref{eqn:CTAnodek} because the former shares updated weight estimates in comparison to the latter, and these estimates are expected to be less noisy; see the analysis further ahead and also \cite{Cattivelli10TSP,Sayed13Chapter}.

An important factor affecting the mean-square performance of diffusion LMS algorithms is the choice of the combination coefficients $\alpha$ and $\beta$. Different combination rules have been proposed in the literature, such as uniform, Laplacian, maximum degree, Metropolis, relative degree, relative degree-variance, and relative variance (which were listed in Table III of reference \cite{Cattivelli10TSP}; see also \cite{Tu11CAMSAP,Sayed13Chapter}). Apart from these static combination rules, where the coefficients are kept constant over time, adaptive rules are also possible. In the adaptive case, the combination weights can be adjusted regularly so that the network can respond to real-time node conditions \cite{Lopes08TSP,Takahashi10TSP,Tu11CAMSAP,Zhao12TSP,Sayed13Chapter}.

Now that we have introduced the various strategies (non-cooperative LMS, block LMS, incremental LMS, ATC and CTA diffusion LMS), we proceed to derive expressions for the \emph{optimal} mean-square performance of the diffusion algorithms \eqref{eqn:CTAnodek} and \eqref{eqn:ATCnodek}. The analysis will highlight some useful properties for distributed algorithms in comparison to centralized counterparts. For example, the results will establish that the diffusion strategies using optimized combination weights perform better than the centralized solutions \eqref{eqn:blockLMS} and \eqref{eqn:incrementalLMS}. Obviously, by assuming knowledge of the network topology, a fusion center can implement the optimized diffusion strategies centrally and therefore attain the same performance as the distributed solution. We are not interested in such situations where the fusion center implements solutions that are fundamentally distributed in nature. We are instead interested in comparing truly distributed solutions of the diffusion type \eqref{eqn:CTAnodek} and \eqref{eqn:ATCnodek} with traditional centralized solutions of the block and incremental LMS types \eqref{eqn:blockLMS} and \eqref{eqn:incrementalLMS}; all with similar levels of LMS complexity.

\section{Performance Analysis for Two-Node Adaptive Networks}
We rely on the energy conservation arguments \cite{Sayed08} to conduct the mean-square performance analysis of two-node LMS adaptive networks. We first compute the individual and network EMSE and MSD for the CTA and ATC algorithms \eqref{eqn:CTAnodek} and \eqref{eqn:ATCnodek}, and then deal with the block and incremental algorithms \eqref{eqn:blockLMS} and \eqref{eqn:incrementalLMS}. The analysis in the sequel is carried out under Assumptions \ref{asm:all}--\ref{asm:uniform} and condition \eqref{eqn:stepsize}. Assumption \ref{asm:smallstepsize} helps ensure the mean-square convergence of the various adaptive strategies that we are considering here --- see, e.g., \cite{Sayed08,Lopes08TSP,Cattivelli10TSP,Sayed13Chapter}. By mean-square convergence of the distributed and centralized algorithms, we mean that $\E{\wt{\bm{w}}}_{k,i}\rightarrow0$, $\E{\wt{\bm{w}}}_{i}\rightarrow0$, and $\E\|{\wt{\bm{w}}}_{k,i}\|^2$ and $\E\|{\wt{\bm{w}}}_{i}\|^2$ tend to constant bounded values as $i\rightarrow\infty$. In addition, Assumption \ref{asm:uniform} and condition \eqref{eqn:stepsize} will help enforce similar convergence rates for all strategies.

\subsection{EMSE and MSD for Non-Cooperative Nodes}
Under Assumptions \ref{asm:all}--\ref{asm:uniform} and as mentioned before, it is known that the EMSE and MSD of the two stand-alone LMS filters in \eqref{eqn:standaloneLMS}, which operate independently of each other, are given by
\begin{equation}
\boxed{
\label{eqn:EMSEindnode}
{\EMSE}_{\ind,k}\approx\frac{\mu\sigma_{v,k}^2\Tr(R_u)}{2}
}\end{equation}
and
\begin{equation}
\boxed{
\label{eqn:MSDindnode}
{\MSD}_{\ind,k}\approx\frac{\mu\sigma_{v,k}^2M}{2}
}\end{equation}
for $k=1,2$. Using \eqref{eqn:EMSEindnode}, the average EMSE and MSD of both nodes are
\begin{equation}
\boxed{
\label{eqn:EMSEindlms}
{\ol{\EMSE}}_{\ind}\approx\frac{\mu\Tr(R_u)}{2}\frac{\sigma_{v,1}^2+\sigma_{v,2}^2}{2}
}\end{equation}
and
\begin{equation}
\boxed{
\label{eqn:MSDindlms}
{\ol{\MSD}}_{\ind}\approx\frac{\mu M}{2}\frac{\sigma_{v,1}^2+\sigma_{v,2}^2}{2}
}\end{equation}

\subsection{EMSE and MSD for Diffusion Algorithms}
Rather than study CTA and ATC algorithms separately, we follow the analysis in \cite{Cattivelli10TSP,Sayed13Chapter} and consider a more general algorithm structure that includes CTA and ATC as special cases. We derive expressions for the node EMSE and MSD for the general structure and then specialize the results for CTA and ATC. Thus, consider a diffusion strategy of the following general form:
\begin{align}
\label{eqn:idealDiffusionPriorDiff}
{\bs{\phi}}_{k,i-1}&=p_{1k}{\bs{w}}_{1,i-1}+p_{2k}{\bs{w}}_{2,i-1}\\
\label{eqn:idealDiffusionIncremental}
{\bs{\psi}}_{k,i}&={\bs{\phi}}_{k,i-1}+\mu{\bs{u}}_{k,i}^*\left[{\bs{d}}_{k}(i)-{\bs{u}}_{k,i}{\bs{\phi}}_{k,i-1}\right]\\
\label{eqn:idealDiffusionPostDiff}
{\bs{w}}_{k,i}&=q_{1k}{\bs{\psi}}_{1,i}+q_{2k}{\bs{\psi}}_{2,i}
\end{align}
where $\{p_{lk},q_{lk}\}$ are the nonnegative entries of $2\times2$ matrices $\{P,Q\}$. The CTA algorithm \eqref{eqn:CTAnodek} corresponds to the special choice $P=A$ and $Q=I_2$ while the ATC algorithm \eqref{eqn:ATCnodek} corresponds to the special choice $P=I_2$ and $Q=A$, where $I_2$ denotes the $2\times2$ identity matrix. From \eqref{eqn:A2def}, it can be verified that the eigenvalues of $A$ are $\{1,\alpha+\beta-1\}$. In the cooperative case, we rule out the choice $\alpha=\beta=1$ so that the two eigenvalues of $A$ are distinct and, hence, $A$ is diagonalizable. Let $A=TDT^{-1}$ denote the eigen-decomposition of $A$:
\begin{align}
\label{eqn:W2Teigdecomp}
&\underbrace{\begin{bmatrix}
  \alpha \;&\; 1-\beta \\
  1-\alpha \;&\; \beta
\end{bmatrix}}_{A}=\underbrace{\frac{1}{2-\alpha-\beta}\begin{bmatrix}
  1-\beta \;&\; 1 \\
  1-\alpha \;&\; -1
\end{bmatrix}}_{T}\cdot\underbrace{\begin{bmatrix}
  1 \;&\; 0 \\
  0 \;&\; \alpha+\beta-1
\end{bmatrix}}_{D}\cdot\underbrace{\begin{bmatrix}
  1 \;&\; 1 \\
  1-\alpha \;&\; \beta-1
\end{bmatrix}}_{T^{-1}}
\end{align}
and let $\lambda_m$ denote the $m$th eigenvalue of $R_u$ whose size is $M\times M$. We derive in Appendix \ref{app:DerivationEMSE} the following expression for the EMSE at node $k$:
\begin{align}
\label{eqn:diffEMSEdefnew}
{\EMSE}_{{\textrm{diff}},k}&\approx\mu^2\sum_{m=1}^{M}\lambda_m^2{\mathrm{vec}}(T^{\T}Q^{\T}R_vQT)^{\T}
(I_4-\xi_mD\otimes D)^{-1}{\mathrm{vec}}(T^{-1}E_{kk}T^{-\T})
\end{align}
for $k=1,2$, where
\begin{align}
\label{eqn:ximdef}
\xi_m\defeq1-2\mu\lambda_m,\qquad R_v\defeq\diag\{\sigma_{v,1}^2,\sigma_{v,2}^2\}
\end{align}
and $E_{kk}$ are $2\times2$ matrices that are given by
\begin{align}
E_{11}=\diag\{1,0\},\qquad E_{22}=\diag\{0,1\}
\end{align}
Likewise, we can derive the MSD at node $k$:
\begin{align}
\label{eqn:diffMSDdef}
{\MSD}_{{\textrm{diff}},k}&\approx\mu^2\sum_{m=1}^{M}\lambda_m{\mathrm{vec}}(T^{\T}Q^{\T}R_vQT)^{\T}
(I_4-\xi_mD\otimes D)^{-1}{\mathrm{vec}}(T^{-1}E_{kk}T^{-\T})
\end{align}
for $k=1,2$. Comparing \eqref{eqn:diffEMSEdefnew} and \eqref{eqn:diffMSDdef} we note that $\lambda_m^2$ in \eqref{eqn:diffEMSEdefnew} is replaced by $\lambda_m$ in \eqref{eqn:diffMSDdef}; all the other factors are identical.

\subsection{EMSE and MSD of CTA Diffusion LMS}
Setting $Q=I_2$, we specialize \eqref{eqn:diffEMSEdefnew} to obtain the EMSE expression for the CTA algorithm:
\begin{align}
\label{eqn:EMSEctaext}
{\EMSE}_{\mathrm{cta},k}&\approx\mu^2\sum_{m=1}^{M}\lambda_m^2{\mathrm{vec}}(T^{\T}R_vT)^{\T}
(I_4-\xi_mD\otimes D)^{-1}{\mathrm{vec}}(T^{-1}E_{kk}T^{-\T})
\end{align}
for $k=1,2$. Substituting \eqref{eqn:W2Teigdecomp} into \eqref{eqn:EMSEctaext}, some algebra will show that the network EMSE for the CTA algorithm, which is defined as the average EMSE of the individual nodes, is given by
\begin{align}
\label{eqn:EMSEctanetwork}
\ol{{\EMSE}}_{\cta}&\approx\sum_{m=1}^{M}\frac{\mu^2\lambda_m^2}{(2 - \alpha - \beta)^2}
\left[\frac{\sigma_{v,1}^2(1 - \beta)^2+\sigma_{v,2}^2(1 - \alpha)^2}{1 - \xi_m}
+\frac{(\alpha-\beta)[\sigma_{v,2}^2(1-\alpha)-\sigma_{v,1}^2(1-\beta)]}{1-\xi_m(\alpha+\beta-1)}\right.\nonumber\\
{}&\qquad\qquad\qquad\qquad\qquad\left.+\frac{(\sigma_{v,1}^2+\sigma_{v,2}^2)[(1-\alpha)^2+(1-\beta)^2]}{2[1-\xi_m(\alpha+\beta-1)^2]}\right]
\end{align}
We argue in Appendix \ref{app:MinimizeEMSEcta} that, under Assumption \ref{asm:smallstepsize} (i.e., for sufficiently small step-sizes), the network EMSE in \eqref{eqn:EMSEctanetwork} is essentially minimized when $\{\alpha,\beta\}$ are chosen as
\begin{equation}
\boxed{
\label{eqn:OptCondCTA}
\alpha=\frac{\sigma_{v,1}^{-2}}{\sigma_{v,1}^{-2}+\sigma_{v,2}^{-2}},\qquad
\beta=\frac{\sigma_{v,2}^{-2}}{\sigma_{v,1}^{-2}+\sigma_{v,2}^{-2}}
}\end{equation}
This choice coincides with the relative degree-variance rule proposed in \cite{Cattivelli10TSP}.\footnote{There is a typo in Table III of \cite{Cattivelli10TSP}, where the noise variances for the relative degree-variance rule should appear inverted.} In the sequel we will compare the performance of the diffusion strategies that result from this choice of combination weights against the performance of the block and incremental LMS strategies \eqref{eqn:blockLMS} and \eqref{eqn:incrementalLMS}.

The value of \eqref{eqn:EMSEctanetwork} that corresponds to the choice \eqref{eqn:OptCondCTA} is then given by
\begin{equation}
\boxed{
\label{eqn:EMSEctaoptnetwork}
{\ol{\EMSE}}_{\cta}^{\opt} \approx \frac{\sigma_{v,1}^2\sigma_{v,2}^2}{\sigma_{v,1}^2 + \sigma_{v,2}^2}
\frac{\mu\Tr(R_u)}{2} + \frac{\mu^2(\sigma_{v,1}^4 + \sigma_{v,2}^4)}{2\,(\sigma_{v,1}^2 + \sigma_{v,2}^2)}
\sum_{m=1}^{M}\lambda_m^2
}\end{equation}
and the corresponding EMSE values at the nodes are
\begin{equation}
\boxed{
\label{eqn:EMSEctafinal}
{\EMSE}_{\cta,k}^{\opt}\approx\frac{\sigma_{v,1}^2\sigma_{v,2}^2}{\sigma_{v,1}^2 + \sigma_{v,2}^2}
\frac{\mu\Tr(R_u)}{2} + \frac{\mu^2\sigma_{v,k}^4}{\sigma_{v,1}^2 + \sigma_{v,2}^2}\sum_{m=1}^{M}\lambda_m^2
}\end{equation}
for $k=1,2$. Similarly, the network MSD is approximately minimized for the same choice \eqref{eqn:OptCondCTA} and its value is given by
\begin{equation}
\boxed{
\label{eqn:MSDctaoptnetwork}
{\ol{\MSD}}_{\cta}^{\opt}\approx\frac{\sigma_{v,1}^2\sigma_{v,2}^2}{\sigma_{v,1}^2 + \sigma_{v,2}^2}
\frac{\mu M}{2} + \frac{\sigma_{v,1}^4 + \sigma_{v,2}^4}{\sigma_{v,1}^2 + \sigma_{v,2}^2}\frac{\mu^2\Tr(R_u)}{2}
}\end{equation}
The corresponding MSD values at the nodes are
\begin{equation}
\boxed{
\label{eqn:MSDctafinal}
{\MSD}_{\cta,k}^{\opt}\approx\frac{\sigma_{v,1}^2\sigma_{v,2}^2}{\sigma_{v,1}^2 + \sigma_{v,2}^2}
\frac{\mu M}{2} + \frac{\sigma_{v,k}^4}{\sigma_{v,1}^2 + \sigma_{v,2}^2}\mu^2\Tr(R_u)
}\end{equation}
for $k=1,2$. We shall refer to the CTA diffusion algorithm that uses \eqref{eqn:OptCondCTA} as the optimal CTA implementation. Note that selecting the coefficients as in \eqref{eqn:OptCondCTA} requires knowledge of the noise variances at both nodes.
This information is usually unavailable. Nevertheless, it is possible to develop adaptive strategies to adjust the coefficients $\{\alpha,\beta\}$ on the fly based on the available data without requiring the nodes to know beforehand the noise profile in the network (see \cite{Tu11CAMSAP,Zhao12TSP,Sayed13Chapter} and \eqref{eqn:MHrule} and \eqref{eqn:estimatenoisevariance} further ahead). We therefore continue the analysis by assuming the nodes are able to determine (or learn) the coefficients \eqref{eqn:OptCondCTA}.

\subsection{EMSE and MSD of ATC Diffusion LMS}
Likewise, setting $Q=A$, we specialize \eqref{eqn:diffEMSEdefnew} to obtain the EMSE expression for the ATC algorithm:
\begin{align}
\label{eqn:EMSEatcext}
{\EMSE}_{\mathrm{atc},k}&\approx\mu^2\sum_{m=1}^{M}\lambda_m^2{\mathrm{vec}}(T^{\T}A^{\T}R_vAT)^{\T}
(I_4 - \xi_mD\otimes D)^{-1}{\mathrm{vec}}(T^{-1}E_{kk}T^{-\T})
\end{align}
for $k=1,2$. Following similar arguments to the CTA case, the network EMSE is given by
\begin{align}
\label{eqn:EMSEatcnetwork}
{}&\ol{{\EMSE}}_{\atc}\approx\sum_{m=1}^{M}\frac{\mu^2\lambda_m}{(2 - \alpha - \beta)^2}
\left[\frac{\sigma_{v,1}^2(1 - \beta)^2 + \sigma_{v,2}^2(1 - \alpha)^2}{1 - \xi_m}\right.\nonumber\\
{}&\qquad\qquad\qquad\qquad\qquad\qquad\quad+\frac{(\alpha - \beta)(\alpha + \beta - 1)[\sigma_{v,2}^2(1 - \alpha) - \sigma_{v,1}^2(1 - \beta)]}
{1 - \xi_m(\alpha + \beta - 1)}\nonumber\\
{}&\qquad\qquad\qquad\qquad\qquad\qquad\quad\left.+\frac{(\sigma_{v,1}^2 + \sigma_{v,2}^2)
(\alpha + \beta - 1)^2[(1 - \alpha)^2 + (1 - \beta)^2]}{2\,[1 - \xi_m(\alpha + \beta-1)^2]}\right]
\end{align}
We can again verify that, under Assumption \ref{asm:smallstepsize}, expression \eqref{eqn:EMSEatcnetwork} is approximately minimized for the same choice \eqref{eqn:OptCondCTA} (see Appendix \ref{app:MinimizeEMSEcta}). The resulting network EMSE value is given by
\begin{equation}
\boxed{
\label{eqn:EMSEatcoptnetwork}
{\ol{\EMSE}}_{\atc}^{\opt}\approx\frac{\sigma_{v,1}^2\sigma_{v,2}^2}
{\sigma_{v,1}^2+\sigma_{v,2}^2}\frac{\mu\Tr(R_u)}{2}
}\end{equation}
and the corresponding EMSE values at the nodes are
\begin{equation}
\boxed{
\label{eqn:EMSEatcfinal}
{\EMSE}_{\atc,k}^{\opt}\approx\frac{\sigma_{v,1}^2\sigma_{v,2}^2}{\sigma_{v,1}^2+\sigma_{v,2}^2}
\frac{\mu\Tr(R_u)}{2}
}\end{equation}
for $k=1,2$. Similarly, the network MSD is approximately minimized for the same choice \eqref{eqn:OptCondCTA}; its value is given by
\begin{equation}
\boxed{
\label{eqn:MSDatcoptnetwork}
{\ol{\MSD}}_{\atc}^{\opt}\approx\frac{\sigma_{v,1}^2\sigma_{v,2}^2}
{\sigma_{v,1}^2+\sigma_{v,2}^2}\frac{\mu M}{2}
}\end{equation}
and the corresponding MSD values at the nodes are
\begin{equation}
\boxed{
\label{eqn:MSDatcfinal}
{\MSD}_{\atc,k}^{\opt}\approx\frac{\sigma_{v,1}^2\sigma_{v,2}^2}{\sigma_{v,1}^2+\sigma_{v,2}^2}\frac{\mu M}{2}
}\end{equation}
for $k=1,2$. We shall refer to the ATC diffusion algorithm that uses \eqref{eqn:OptCondCTA} as the optimal ATC implementation.

\subsection{Uniform CTA and ATC Diffusion LMS}
Uniform CTA and ATC diffusion LMS correspond to the choice $\alpha=\beta=0.5$, which means that the two nodes equally trust each other's estimates. This situation coincides with the uniform combination rule \cite{Cattivelli10TSP}. According to \eqref{eqn:EMSEctanetwork} and \eqref{eqn:diffMSDdef}, the network EMSE and MSD for uniform CTA are
\begin{equation}
\boxed{
\label{eqn:EMSEequalCTAnetwork}
{\ol{\EMSE}}_{\cta}^{\unf}\approx\frac{\sigma_{v,1}^2
+\sigma_{v,2}^2}{4}\left(\frac{\mu\Tr(R_u)}{2}+\mu^2\sum_{m=1}^{M}\lambda_m^2\right)
}\end{equation}
and
\begin{equation}
\boxed{
\label{eqn:MSDequalCTAnetwork}
{\ol{\MSD}}_{\cta}^{\unf}\approx\frac{\sigma_{v,1}^2
+\sigma_{v,2}^2}{4}\left(\frac{\mu M}{2}+\mu^2\Tr(R_u)\right)
}\end{equation}
Similarly, according to \eqref{eqn:EMSEatcnetwork} and \eqref{eqn:diffMSDdef}, the network EMSE and MSD for uniform ATC are
\begin{equation}
\boxed{
\label{eqn:EMSEequalATCnetwork}
{\ol{\EMSE}}_{\atc}^{\unf}\approx\frac{\sigma_{v,1}^2
+\sigma_{v,2}^2}{4}\frac{\mu\Tr(R_u)}{2}
}\end{equation}
and
\begin{equation}
\boxed{
\label{eqn:MSDequalATCnetwork}
{\ol{\MSD}}_{\atc}^{\unf}\approx\frac{\sigma_{v,1}^2
+\sigma_{v,2}^2}{4}\frac{\mu M}{2}
}\end{equation}

\subsection{EMSE and MSD of Block LMS and Incremental LMS}
In Appendix \ref{app:DerivationEMSEvec}, we derive the EMSE and MSD for the block LMS implementation \eqref{eqn:blockLMS} and arrive at
\begin{equation}
\label{eqn:EMSEveclms}
\boxed{
{\EMSE}_{\blk}\approx\frac{\sigma_{v,1}^2+\sigma_{v,2}^2}{2}\frac{\mu'\Tr(R_u)}{2}}
\end{equation}
and
\begin{equation}
\label{eqn:MSDveclms}
\boxed{
{\MSD}_{\blk}\approx\frac{\sigma_{v,1}^2+\sigma_{v,2}^2}{2}\frac{\mu'M}{2}}
\end{equation}
With regards to the incremental LMS algorithm \eqref{eqn:incrementalLMS}, we note from Assumption \ref{asm:smallstepsize} that the step-size $\mu'$ is sufficiently small so that we can assume $\mu'\Tr(R_u)\ll1$. Then, from \eqref{eqn:incrementalLMS} we get
\begin{align}
\label{eqn:incremental2}
{\bs{w}}_{i}&={\bs{w}}_{i-1}+\mu'[{\bs{u}}_{1,i}^*({\bs{d}}_{1}(i)-{\bs{u}}_{1,i}{\bs{w}}_{i-1})
+{\bs{u}}_{2,i}^*({\bs{d}}_{2}(i)-{\bs{u}}_{2,i}{\bs{w}}_{i-1})]
+\underbrace{\mu'^2\|{\bs{u}}_{2,i}\|^2{\bs{u}}_{1,i}^*({\bs{d}}_{1}(i)-{\bs{u}}_{1,i}{\bs{w}}_{i-1})}_{O(\mu'^2)}\nonumber\\
{}&\approx{\bs{w}}_{i-1}+\mu'\begin{bmatrix}
{\bs{u}}_{1,i} \\
{\bs{u}}_{2,i} \\
\end{bmatrix}^*\left(\begin{bmatrix}
{\bs{d}}_{1}(i) \\
{\bs{d}}_{2}(i) \\
\end{bmatrix}-\begin{bmatrix}
{\bs{u}}_{1,i} \\
{\bs{u}}_{2,i} \\
\end{bmatrix}{\bs{w}}_{i-1}\right)
\end{align}
which means that the incremental LMS update \eqref{eqn:incrementalLMS} can be well approximated by the block LMS update \eqref{eqn:blockLMS}. Then, the EMSE of incremental LMS \eqref{eqn:incrementalLMS} can be well approximated by (reference \cite{Lopes07TSP} provides a more detailed analysis of the performance of adaptive incremental LMS strategies):
\begin{equation}
\label{eqn:EMSEinclms}
\boxed{
{\EMSE}_{\inc}\approx\frac{\sigma_{v,1}^2+\sigma_{v,2}^2}{2}\frac{\mu'\Tr(R_u)}{2}}
\end{equation}
and its MSD as
\begin{equation}
\label{eqn:MSDinclms}
\boxed{
{\MSD}_{\inc}\approx\frac{\sigma_{v,1}^2+\sigma_{v,2}^2}{2}\frac{\mu'M}{2}}
\end{equation}
It is worth noting that although \eqref{eqn:EMSEveclms} and \eqref{eqn:EMSEinclms} are similar for small step-sizes, incremental LMS actually outperforms block LMS \cite{Cattivelli11TSPincremental} because the former uses the intermediate estimate ${\bs{\phi}}_{i}$ during one step of the update in \eqref{eqn:incrementalLMS} while the latter does not. The intermediate estimate ${\bs{\phi}}_{i}$ is generally ``less noisy'' than ${\bs{w}}_{i-1}$ so that incremental LMS generally outperforms block LMS. However, we shall not distinguish between incremental LMS and block LMS in this work, when we compare their performance with other strategies in the sequel.

\begin{table*}[t!]
  \caption{Network EMSE and MSD for Various Strategies over Two-Node LMS Adaptive Networks}
  \label{table1}
  \centering
  \begin{threeparttable}
  \renewcommand{\arraystretch}{1.5}
  \begin{tabular}{|c|cr|cr|}
  \hline
  \bfseries Type & \bfseries \qquad Network EMSE & {} & \bfseries \qquad Network MSD & {}\\
  \hline
  Optimal ATC  \eqref{eqn:ATCnodek}& $c_1{\sigma}_{\textrm{harm}}^{2}$ & \eqref{eqn:EMSEatcoptnetwork} & $c_1'{\sigma}_{\textrm{harm}}^{2}$ & \eqref{eqn:MSDatcoptnetwork}\\
  \hline
  Optimal CTA  \eqref{eqn:CTAnodek}& $c_1{\sigma}_{\textrm{harm}}^{2}+c_2\left(2{\sigma}_{\textrm{arth}}^{2}-
  {\sigma}_{\textrm{harm}}^{2}\right)$ & \eqref{eqn:EMSEctaoptnetwork} & $c_1'{\sigma}_{\textrm{harm}}^{2}+c_2'\left(2{\sigma}_{\textrm{arth}}^{2}-
  {\sigma}_{\textrm{harm}}^{2}\right)$ & \eqref{eqn:MSDctaoptnetwork}\\
  \hline
  Uniform ATC  \eqref{eqn:ATCnodek}& $c_1{\sigma}_{\textrm{arth}}^{2}$ & \eqref{eqn:EMSEequalATCnetwork} & $c_1'{\sigma}_{\textrm{arth}}^{2}$ & \eqref{eqn:MSDequalATCnetwork}\\
  \hline
  Uniform CTA  \eqref{eqn:CTAnodek}& $\left(c_1+c_2\right){\sigma}_{\textrm{arth}}^{2}$ & \eqref{eqn:EMSEequalCTAnetwork} & $\left(c_1'+c_2'\right){\sigma}_{\textrm{arth}}^{2}$ & \eqref{eqn:MSDequalCTAnetwork}\\
  \hline
  Block LMS  \eqref{eqn:blockLMS}& $2c_3{\sigma}_{\textrm{arth}}^{2}$ & \eqref{eqn:EMSEveclms} & $2c_3'{\sigma}_{\textrm{arth}}^{2}$ & \eqref{eqn:MSDveclms}\\
  \hline
  Incremental LMS  \eqref{eqn:incrementalLMS}& $2c_3{\sigma}_{\textrm{arth}}^{2}$ & \eqref{eqn:EMSEinclms} & $2c_3'{\sigma}_{\textrm{arth}}^{2}$ & \eqref{eqn:MSDinclms}\\
  \hline
  Stand-alone LMS  \eqref{eqn:standaloneLMS}& $2c_1{\sigma}_{\textrm{arth}}^{2}$ & \eqref{eqn:EMSEindlms} & $2c_1'{\sigma}_{\textrm{arth}}^{2}$ & \eqref{eqn:MSDindlms}\\
  \hline
  \end{tabular}
  \smallvskip
  \begin{tablenotes}
    \item[1] $\lambda\defeq{\col}\{\lambda_1,\dots,\lambda_N\}$ consists of the eigenvalues of $R_u$, ${\sigma}_{\textrm{arth}}^{2}\defeq\frac{\sigma_{v,1}^2+\sigma_{v,2}^2}{2}$ and ${\sigma}_{\textrm{harm}}^{2}\defeq\frac{2\sigma_{v,1}^2\sigma_{v,2}^2}{\sigma_{v,1}^2+\sigma_{v,2}^2}$.
    \item[2] $c_1\defeq\frac{\mu\Tr(R_u)}{4}$, $c_2\defeq\frac{\mu^2\|\lambda\|^2}{2}$, $c_3\defeq\frac{\mu'\Tr(R_u)}{4}$, $c_1'\defeq\frac{\mu M}{4}$, $c_2'\defeq\frac{\mu^2\Tr(R_u)}{2}$, and $c_3'\defeq\frac{\mu'M}{4}$.
  \end{tablenotes}
  \end{threeparttable}
\end{table*}

\begin{table*}[t!]
  \caption{EMSE for the Individual Nodes in Various Strategies over Two-Node LMS Adaptive Networks}
  \label{table2}
  \centering
  \renewcommand{\arraystretch}{1.5}
  \begin{tabular}{|c|cr|cr|}
  \hline
  \bfseries Type & \bfseries \qquad EMSE of Node 1 & {} & \bfseries \qquad EMSE of Node 2 & {} \\
  \hline
  Optimal ATC \eqref{eqn:ATCnodek} &
  $c_1{\sigma}_{\textrm{harm}}^{2}$ & \eqref{eqn:EMSEatcfinal} & $c_1{\sigma}_{\textrm{harm}}^{2}$ & \eqref{eqn:EMSEatcfinal} \\
  \hline
  Optimal CTA \eqref{eqn:CTAnodek} &
  $c_1{\sigma}_{\textrm{harm}}^{2}+c_2\frac{\sigma_{v,1}^4}{{\sigma}_{\textrm{arth}}^{2}}$ & \eqref{eqn:EMSEctafinal} &
  $c_1{\sigma}_{\textrm{harm}}^{2}+c_2\frac{\sigma_{v,2}^4}{{\sigma}_{\textrm{arth}}^{2}}$ & \eqref{eqn:EMSEctafinal} \\
  \hline
  Stand-alone LMS \eqref{eqn:standaloneLMS} & $2c_1\sigma_{v,1}^2$ & \eqref{eqn:EMSEindnode} & $2c_1\sigma_{v,2}^2$ & \eqref{eqn:EMSEindnode} \\
  \hline
  \end{tabular}
\end{table*}

\begin{table*}[t!]
  \caption{MSD for the Individual Nodes in Various Strategies over Two-Node LMS Adaptive Networks}
  \label{table3}
  \centering
  \renewcommand{\arraystretch}{1.5}
  \begin{tabular}{|c|cr|cr|}
  \hline
  \bfseries Type & \bfseries \qquad MSD of Node 1 & {} & \bfseries \qquad MSD of Node 2 & {} \\
  \hline
  Optimal ATC \eqref{eqn:ATCnodek} &
  $c_1'{\sigma}_{\textrm{harm}}^{2}$ & \eqref{eqn:MSDatcfinal} & $c_1'{\sigma}_{\textrm{harm}}^{2}$ & \eqref{eqn:MSDatcfinal} \\
  \hline
  Optimal CTA \eqref{eqn:CTAnodek} &
  $c_1'{\sigma}_{\textrm{harm}}^{2}+c_2'\frac{\sigma_{v,1}^4}{{\sigma}_{\textrm{arth}}^{2}}$ & \eqref{eqn:MSDctafinal}&
  $c_1'{\sigma}_{\textrm{harm}}^{2}+c_2'\frac{\sigma_{v,2}^4}{{\sigma}_{\textrm{arth}}^{2}}$ & \eqref{eqn:MSDctafinal} \\
  \hline
  Stand-alone LMS \eqref{eqn:standaloneLMS} & $2c_1'\sigma_{v,1}^2$ & \eqref{eqn:MSDindnode} & $2c_1'\sigma_{v,2}^2$ & \eqref{eqn:MSDindnode} \\
  \hline
  \end{tabular}
\end{table*}

\subsection{Summary}
We list the expressions for the network EMSE and MSD for the various strategies under Assumptions \ref{asm:all}--\ref{asm:uniform} in Table \ref{table1}, and the expressions for the individual nodes in Tables \ref{table2} and \ref{table3}, respectively. It is worth noting from these expressions that the EMSE is dependent on the step-size parameter. In order to compare the EMSE of the algorithms in a fair manner, the step-sizes need to be tuned appropriately because algorithms generally differ in terms of their convergence rates and steady-state performance. Some algorithms converge faster but may have larger EMSE. Others may have smaller EMSE but converge slower. Therefore, the step-sizes should be adjusted in such a way that all algorithms exhibit similar convergence rates. Then, under these conditions, the EMSE values can be fairly compared. We proceed to explain this issue in greater detail in the next section.

\section{Performance Comparison for Various Adaptive Strategies}
Adaptive algorithms differ in their mean-square convergence rates and in their steady-state mean-square error performance. In order to ensure a fair comparison among algorithms, we should either fix their convergence rates at the same value and then compare the resulting  steady-state performance, or we should fix the steady-state performance and then compare the convergence rates. To clarify this procedure further, we consider the concept of ``operation curves (OC)''.

\subsection{Operation Curves for Adaptive Strategies}
The OC of an algorithm has two axes: the horizontal axis represents its EMSE and the vertical axis represents its (mean-square) convergence rate. Each point on the OC corresponds to a choice of the step-size parameter. Now the EMSE and convergence rate of an adaptive implementation, such as stand-alone LMS, are both dependent on the step-size parameter used by the algorithm. For example, under Assumptions \ref{asm:all}--\ref{asm:uniform}, the EMSE of a stand-alone LMS filter of the type \eqref{eqn:standaloneLMS}, denoted by $\zeta(\mu)$, is a function of $\mu$ and is given by \cite{Sayed08}:
\begin{align}
\label{eqn:EMSEstandalone}
\zeta(\mu)&\approx\frac{\mu\sigma_{v}^2\Tr(R_u)}{2}
\end{align}
The function $\zeta(\mu)$ is monotonically increasing in $\mu$. It is clear from \eqref{eqn:EMSEstandalone} that the smaller the value of $\mu$, the lower the EMSE (which is desirable). However, a smaller step-size $\mu$ results in slower mean-square convergence. This is because, under Assumptions \ref{asm:all}--\ref{asm:uniform}, the modes of convergence for a stand-alone LMS implementation \eqref{eqn:standaloneLMS} are approximately given by \cite[p.\,360]{Sayed08}:
\begin{align}
\label{eqn:eigOmegaapprox}
\xi_m\defeq1-2\mu\lambda_m
\end{align}
for $m=1,\dots,M$, where the $\{\lambda_m\}$ are the eigenvalues of $R_u$. The value of $\xi_m$ that is closest to the unit circle determines the rate of convergence of $\E{\wt{\bm{w}}}_i$ and $\E\|{\wt{\bm{w}}}_i\|^2$ towards their steady-state values. It is clear from \eqref{eqn:eigOmegaapprox} that the smaller $\mu$ is, the closer the mode is to the unit circle, and the slower the convergence of the algorithm will be. Hence, under Assumption \ref{asm:smallstepsize},
\begin{itemize}
  \item Increasing $\mu$ results in faster convergence at the cost of a higher (worse) EMSE.
  \item Decreasing $\mu$ results in slower convergence but a lower (better) EMSE.
\end{itemize}
For this reason, in order to compare fairly the performance of various algorithms, we need to jointly examine their EMSE and convergence rates. It is worth noting that the concept of operation curves can also be applied to other steady-state performance metrics such as the MSD. However, due to space limitations, we focus on the EMSE in this work.

\subsubsection{Operation Curve for Stand-Alone LMS}
For stand-alone LMS filters, under Assumptions \ref{asm:all}--\ref{asm:uniform}, the average network EMSE is given by \eqref{eqn:EMSEindlms} and the dominant mode (the one that is closest to the unit circle) is given by
\begin{align}
\label{eqn:rhoindividual}
{\textrm{mode}}_{\ind}\approx1-2\mu\lambdamin(R_u)
\end{align}
where $\lambdamin(\cdot)$ denotes the smallest eigenvalue of its matrix argument.

\subsubsection{Operation Curve for CTA Diffusion LMS}
Based on Assumptions \ref{asm:all}--\ref{asm:uniform}, the expressions for the network EMSE of optimal CTA and uniform CTA are given by \eqref{eqn:EMSEctaoptnetwork} and \eqref{eqn:EMSEequalCTAnetwork}, respectively. Meanwhile, from expression (43) in \cite{Cattivelli10TSP}, we know that the modes of convergence for CTA algorithms are determined by the eigenvalues of $[A\otimes(I_M-\mu R_u^\T)]\otimes[A\otimes(I_M-\mu R_u)]$. Now recall from \eqref{eqn:W2Teigdecomp} that $A$ has two real eigenvalues at $\{1,\alpha+\beta-1\}$. The second eigenvalue is smaller than one in magnitude. Therefore, the dominant mode for CTA algorithms is given by
\begin{align}
\label{eqn:rhooptcta}
{\textrm{mode}}_{\cta}^{\opt}\approx{\textrm{mode}}_{\cta}^{\unf}\approx1-2\mu\lambdamin(R_u)
\end{align}

\subsubsection{Operation Curve for ATC Diffusion LMS}
Based on Assumptions \ref{asm:all}--\ref{asm:uniform}, the network EMSE for optimal ATC and uniform ATC are given by \eqref{eqn:EMSEatcoptnetwork} and \eqref{eqn:EMSEequalATCnetwork}, respectively. The modes of mean-square convergence for ATC algorithms are also determined by the eigenvalues of $[A\otimes(I_M-\mu R_u^\T)]\otimes[A\otimes(I_M-\mu R_u)]$ \cite{Cattivelli10TSP}. Therefore, the dominant mode for ATC is also
\begin{align}
\label{eqn:rhooptatc}
{\textrm{mode}}_{\atc}^{\opt}\approx{\textrm{mode}}_{\atc}^{\unf}\approx1-2\mu\lambdamin(R_u)
\end{align}

\subsubsection{Operation Curves for Block LMS and Incremental LMS}
The EMSE for block LMS and incremental LMS are given by \eqref{eqn:EMSEveclms} and \eqref{eqn:EMSEinclms}, respectively. In Appendix \ref{app:DerivationEMSEvec}, we show that their dominant mode is
\begin{align}
\label{eqn:rhoblock}
{\textrm{mode}}_{\blk}\approx{\textrm{mode}}_{\inc}\approx1-4\mu'\lambdamin(R_u)
\end{align}

\begin{figure}[t]
  \centering
  \includegraphics[width=3.2in]{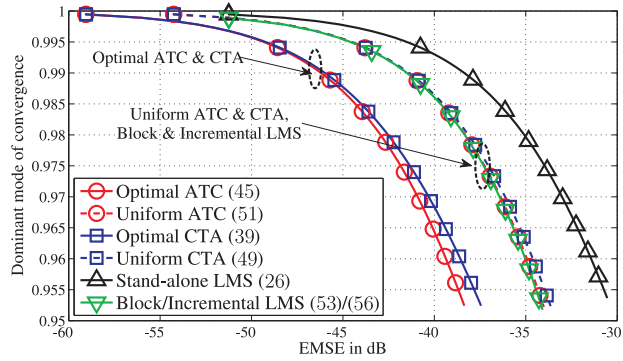}
  \caption{Operation curves for various algorithms when $M=10$, $R_u=I_M$, $\sigma_{v,1}^2=0.01$, and $\sigma_{v,2}^2=0.001$.}
  \label{fig:allOP}
\end{figure}

We plot the operation curves for all algorithms in Fig. \ref{fig:allOP}. From the figure we observe that
(i) optimal ATC and optimal CTA have similar performance and outperform all other strategies; (ii) block LMS and incremental LMS have similar performance to uniform ATC and uniform CTA; (iii) the non-cooperative stand-alone LMS implementation has the worst performance. In the following, we shall fix the convergence rate at the same value for all strategies and then compare their EMSE levels analytically.

\subsection{Common Convergence Rate}
As was mentioned before, the performance of each algorithm is dictated by two factors: its steady-state EMSE and its mean-square convergence rate, and both factors are functions of the step-size $\mu$. In order to make a fair comparison among the algorithms, we shall fix one factor and then compare them in terms of the other factor, and vice versa.

From \eqref{eqn:rhoindividual}, \eqref{eqn:rhooptcta}, and \eqref{eqn:rhooptatc}, we know that ATC algorithms, CTA algorithms, and stand-alone LMS filters have (approximately) the same dominant mode for mean-square convergence:
\begin{align}
\label{eqn:rho1def}
{\textrm{mode}}_1\defeq1-2\mu\lambdamin(R_u)
\end{align}
For block LMS and incremental LMS, from \eqref{eqn:rhoblock}, their dominant mode of convergence is approximately
\begin{align}
\label{eqn:rho2def}
{\textrm{mode}}_2\defeq1-4\mu'\lambdamin(R_u)
\end{align}
In order to make all algorithms converge at the same rate, i.e., ${\textrm{mode}}_1={\textrm{mode}}_2$, we enforce the relation:
\begin{equation}
\boxed{
\label{eqn:mu1andmu2}
\mu=2\mu'
}\end{equation}
An intuitive explanation for \eqref{eqn:mu1andmu2} is that, for a set of data $\{{\bm{d}}_1(i),{\bm{d}}_2(i);{\bm{u}}_{1,i},{\bm{u}}_{2,i}\}$, incremental LMS performs two successive iterations while each stand-alone LMS filter performs only one iteration. For this reason, the step-size of incremental LMS needs to be half the value of that for stand-alone LMS in order for both classes of algorithms to converge at the same rate. Based on condition \eqref{eqn:mu1andmu2}, we now modify Table \ref{table1} into Table \ref{table4}, and proceed to compare the EMSE for various strategies. Although we focus on comparing the EMSE performance, similar arguments can be applied to the MSD performance of the various strategies.

\begin{table}[t!]
  \caption{Network EMSE values from Table \ref{table1} using $\mu=2\mu'$}
  \label{table4}
  \centering
  \begin{threeparttable}
  \renewcommand{\arraystretch}{1.5}
  \begin{tabular}{|c|cr|c|}
  \hline
  \bfseries Type & \bfseries \qquad Network EMSE & {} & \bfseries Acronym \\
  \hline
  Opt. ATC  \eqref{eqn:ATCnodek}& $c_1{\sigma}_{\textrm{harm}}^{2}$ & \eqref{eqn:EMSEatcoptnetwork} & ${\ol{\EMSE}}_{\atc}^{\opt}$ \\
  \hline
  Opt. CTA  \eqref{eqn:CTAnodek}& $c_1{\sigma}_{\textrm{harm}}^{2}+c_2\left(2{\sigma}_{\textrm{arth}}^{2}-
  {\sigma}_{\textrm{harm}}^{2}\right)$ & \eqref{eqn:EMSEctaoptnetwork} & ${\ol{\EMSE}}_{\cta}^{\opt}$ \\
  \hline
  Unf. ATC  \eqref{eqn:ATCnodek}& $c_1{\sigma}_{\textrm{arth}}^{2}$ & \eqref{eqn:EMSEequalATCnetwork} & ${\ol{\EMSE}}_{\atc}^{\unf}$ \\
  \hline
  Unf. CTA  \eqref{eqn:CTAnodek}& $\left(c_1+c_2\right){\sigma}_{\textrm{arth}}^{2}$ & \eqref{eqn:EMSEequalCTAnetwork} & ${\ol{\EMSE}}_{\cta}^{\unf}$ \\
  \hline
  Blk. LMS  \eqref{eqn:blockLMS}& $c_1{\sigma}_{\textrm{arth}}^{2}$ & \eqref{eqn:EMSEveclms} & ${\EMSE}_{\blk}$ \\
  \hline
  Inc. LMS  \eqref{eqn:incrementalLMS}& $c_1{\sigma}_{\textrm{arth}}^{2}$ & \eqref{eqn:EMSEinclms} & ${\EMSE}_{\inc}$ \\
  \hline
  Std. LMS  \eqref{eqn:standaloneLMS}& $2c_1{\sigma}_{\textrm{arth}}^{2}$ & \eqref{eqn:EMSEindlms} & ${\ol{\EMSE}}_{\ind}$ \\
  \hline
  \end{tabular}
  \smallvskip
  \begin{tablenotes}
    \item[1] ${\sigma}_{\textrm{arth}}^{2} \defeq \frac{\sigma_{v,1}^2+\sigma_{v,2}^2}{2}$ and ${\sigma}_{\textrm{harm}}^{2} \defeq \frac{2\sigma_{v,1}^2\sigma_{v,2}^2}{\sigma_{v,1}^2+\sigma_{v,2}^2}$, where $\sigma_{v,2}^2<\sigma_{v,1}^2$.
    \item[2] $c_1\defeq\frac{\mu\Tr(R_u)}{4}$ and $c_2\defeq\frac{\mu^2\|\lambda\|^2}{2}$.
  \end{tablenotes}
  \end{threeparttable}
\end{table}

\begin{table*}[t!]
  \renewcommand{\arraystretch}{1.5}
  \caption{Comparing Network EMSEs for Various Algorithms}
  \label{table5}
  \centering
  \small
  \begin{threeparttable}
  \begin{tabular}{|c|c|c|c|c|c|c|}
  \hline
  {}               & Opt. ATC          & Opt. CTA                    & Unf. ATC                   & Unf. CTA                   & Blk./Inc. LMS\tnote{1}          & Std. LMS        \\
  \hline
  Opt. ATC          & {}            &   \textbf{\emph{better}}\tnote{3}  & \textbf{better}           & \textbf{better}           & \textbf{better}           & \textbf{better} \\
  \hline
  Opt. CTA          & worse\tnote{3}  & {}                      & \textbf{better}\tnote{2} & \textbf{better}           & \textbf{better}\tnote{2} & \textbf{better} \\
  \hline
  Unf. ATC          & worse            & worse\tnote{2}            & {}                     & \textbf{better}\tnote{3} & equal                     & \textbf{better} \\
  \hline
  Unf. CTA          & worse            & worse                      & worse\tnote{3}           & {}                     & worse\tnote{3}           & \textbf{better} \\
  \hline
  Blk./Inc. LMS\tnote{1} & worse            & worse\tnote{2}            & equal                     & \textbf{better}\tnote{3} & {}                     & \textbf{better} \\
  \hline
  Std. LMS         & worse            & worse                      & worse                     & worse                     & worse                     & {}           \\
  \hline
  \end{tabular}
  \smallvskip
  \begin{tablenotes}
    \item[1] The step-sizes for block and incremental LMS are half the value for the other algorithms.
    \item[2] If $2\mu\sigma_u^2<({\sigma}_{\textrm{arth}}^2-{\sigma}_{\textrm{harm}}^2)/(2{\sigma}_{\textrm{arth}}^2-{\sigma}_{\textrm{harm}}^2)$,
        which is generally true under Assumption \ref{asm:smallstepsize}.
    \item[3] By a small margin on the order of $\mu^2$.
  \end{tablenotes}
  \end{threeparttable}
\end{table*}

\subsection{Comparing Network EMSE}
We use Table \ref{table4} to compare the network EMSE. First, the harmonic and arithmetic means of $\{\sigma_{v,1}^2,\sigma_{v,2}^2\}$ are defined as
\begin{align}
{\sigma}_{\textrm{harm}}^{2}\defeq\frac{2\sigma_{v,1}^2\sigma_{v,2}^2}{\sigma_{v,1}^2+\sigma_{v,2}^2},\qquad
{\sigma}_{\textrm{arth}}^{2}\defeq\frac{\sigma_{v,1}^2+\sigma_{v,2}^2}{2}
\end{align}
and it holds that ${\sigma}_{\textrm{harm}}^2<{\sigma}_{\textrm{arth}}^2$. As a result, it is easy to verify that
\begin{equation}
\boxed{
{\ol{\EMSE}}_{\atc}^{\opt}<{\ol{\EMSE}}_{\cta}^{\opt}} \quad \mbox{and} \quad
\boxed{
{\ol{\EMSE}}_{\atc}^{\unf}<{\ol{\EMSE}}_{\cta}^{\unf}
}\end{equation}
although their values are close to each other since $c_2$ is proportional to $\mu^2$ and $\mu$ is assumed to be sufficiently small by Assumption \ref{asm:smallstepsize}. For CTA-type algorithms, it is further straightforward to verify that
\begin{equation}
\boxed{
{\ol{\EMSE}}_{\cta}^{\opt}<{\ol{\EMSE}}_{\cta}^{\unf}
<{\ol{\EMSE}}_{\ind}
}\end{equation}
since, under Assumption \ref{asm:smallstepsize},
\begin{align}
c_2<c_1 & \Longrightarrow
c_2(\sigma_{\textrm{arth}}^2-\sigma_{\textrm{harm}}^2)<c_1(\sigma_{\textrm{arth}}^2-\sigma_{\textrm{harm}}^2) \nonumber\\
& \Longrightarrow
c_1\sigma_{\textrm{harm}}^2+c_2\sigma_{\textrm{arth}}^2<c_1\sigma_{\textrm{arth}}^2+c_2\sigma_{\textrm{harm}}^2 \nonumber\\
& \Longrightarrow
c_1\sigma_{\textrm{harm}}^2+2c_2\sigma_{\textrm{arth}}^2<(c_1+c_2)\sigma_{\textrm{arth}}^2+c_2\sigma_{\textrm{harm}}^2 \nonumber\\
& \Longrightarrow
c_1\sigma_{\textrm{harm}}^2+c_2(2\sigma_{\textrm{arth}}^2-\sigma_{\textrm{harm}}^2)<(c_1+c_2)\sigma_{\textrm{arth}}^2\nonumber
\end{align}
Similarly, for ATC-type algorithms, we get
\begin{equation}
\boxed{
{\ol{\EMSE}}_{\atc}^{\opt}<{\ol{\EMSE}}_{\atc}^{\unf}
<{\ol{\EMSE}}_{\ind}
}\end{equation}
The relation between optimal CTA and uniform ATC depends on the parameters $\{{\sigma}_{\textrm{harm}}^2,{\sigma}_{\textrm{arth}}^2,c_1,c_2\}$ since
\begin{align}
\label{eqn:CmpOptCTAvUnfATC}
{\ol{\EMSE}}_{\cta}^{\opt}<{\ol{\EMSE}}_{\atc}^{\unf}\Longleftrightarrow \frac{c_2}{c_1}<\frac{{\sigma}_{\textrm{arth}}^2-{\sigma}_{\textrm{harm}}^2}{2{\sigma}_{\textrm{arth}}^2-{\sigma}_{\textrm{harm}}^2}
\end{align}
which is usually true under Assumption \ref{asm:smallstepsize}. Uniform ATC, block LMS, and incremental LMS have the same performance:
\begin{equation}
\boxed{
{\EMSE}_{\blk}={\EMSE}_{\inc}={\ol{\EMSE}}_{\atc}^{\unf}
}\end{equation}
Hence, optimal ATC outperforms block LMS and incremental LMS:
\begin{equation}
\boxed{
{\ol{\EMSE}}_{\atc}^{\opt}<{\EMSE}_{\textrm{blk/inc}}
}\end{equation}
The relations between optimal CTA, block LMS, and incremental LMS also depend on the parameters $\{{\sigma}_{\textrm{harm}}^2,{\sigma}_{\textrm{arth}}^2,c_1,c_2\}$ since
\begin{align}
\label{eqn:CmpOptCTAvBlock}
{\ol{\EMSE}}_{\cta}^{\opt}<{\EMSE}_{\textrm{blk/inc}}\Longleftrightarrow
\frac{c_2}{c_1}<\frac{{\sigma}_{\textrm{arth}}^2-{\sigma}_{\textrm{harm}}^2}{2{\sigma}_{\textrm{arth}}^2-{\sigma}_{\textrm{harm}}^2}
\end{align}
which is the same condition as \eqref{eqn:CmpOptCTAvUnfATC}. Block LMS and incremental LMS outperform uniform CTA:
\begin{equation}
\boxed{
{\EMSE}_{\textrm{blk/inc}}<{\ol{\EMSE}}_{\cta}^{\unf}
}\end{equation}
but only by a small margin since $c_2$ is proportional to $\mu^2$. We summarize the network EMSE relationships in Table \ref{table5}. Entries of Table \ref{table5} should be read from left to right. For example, the entry (in italics) on the second row and third column should be read to mean: ``\emph{optimal ATC is better than optimal CTA (i.e., it results in lower EMSE)}''.

\subsection{Comparing Individual Node EMSE}
We compare the EMSE of node 1 under various strategies using Table \ref{table2}. First, node 1 in optimal ATC outperforms that in optimal CTA: ${\EMSE}_{\textrm{atc,1}}^{\opt}<{\EMSE}_{\textrm{cta,1}}^{\opt}$. But more importantly, node 1 in optimal CTA outperforms that in stand-alone LMS because ${\EMSE}_{\textrm{ind,1}}<{\EMSE}_{\textrm{cta,1}}^{\opt}$ when $c_2<c_1$, which is true under Assumption \ref{asm:smallstepsize}. Recall that node 1 has larger noise variance than node 2. Therefore, ATC and CTA cooperation helps it attain better EMSE value than what it would obtain if it operates independently.

Likewise, we compare the EMSE of node 2 using Table \ref{table2}. Node 2 in optimal ATC performs better than that in optimal CTA: ${\EMSE}_{\textrm{atc,2}}^{\opt}<{\EMSE}_{\textrm{cta,2}}^{\opt}$. Again, and importantly, node 2 in optimal CTA outperforms that in stand-alone LMS because ${\EMSE}_{\textrm{cta,2}}^{\opt}<{\EMSE}_{\textrm{ind,2}}$ when $c_2<c_1$, which is again true under Assumption \ref{asm:smallstepsize}. Although node 2 has less noise than node 1, it still benefits from cooperating with node 1 and is able to reduce its EMSE below what it would obtain if it operates independently.

The relations between the EMSE for both nodes 1 and 2 under various strategies are the same --- node 2 always outperforms node 1 due to the lower noise level. Table \ref{table6} summarizes the results.

\begin{table}[t!]
  \renewcommand{\arraystretch}{1.5}
  \caption{Comparing Individual Node EMSE Values for Various Strategies}
  \label{table6}
  \centering
  \begin{tabular}{|c|c|c|c|}
  \hline
  {}      &  Opt. ATC  & Opt. CTA & Std. LMS \\
  \hline
  Opt. ATC & {}    & \textbf{better}  & \textbf{better}  \\
  \hline
  Opt. CTA & worse    & {}   & \textbf{better}  \\
  \hline
  Std. LMS & worse    & worse   & {}   \\
  \hline
  \end{tabular}
\end{table}

\subsection{Simulations Results}
\begin{figure*}
\centering
\includegraphics[width=6.5in]{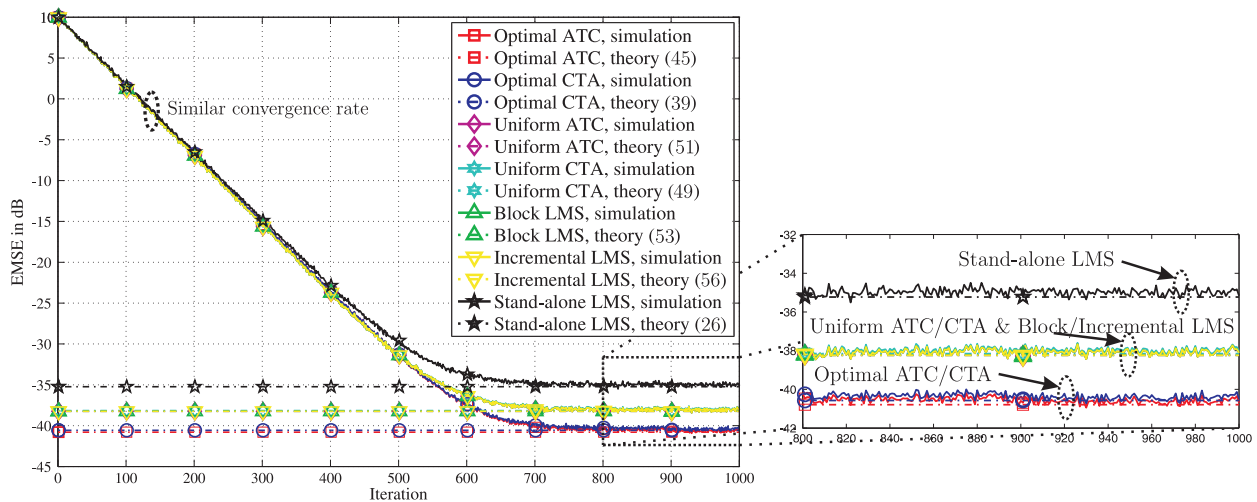}
\caption{Comparison of network EMSE when $M=10$, $R_u=I_M$, $\sigma_{v,1}^2=0.01$, $\sigma_{v,2}^2=0.002$, and $\mu=0.01$.}
\label{fig:network2nodes}
\end{figure*}

\begin{figure*}
\centerline{
\subfloat[Node 1.]
{\includegraphics[width=3.2in]{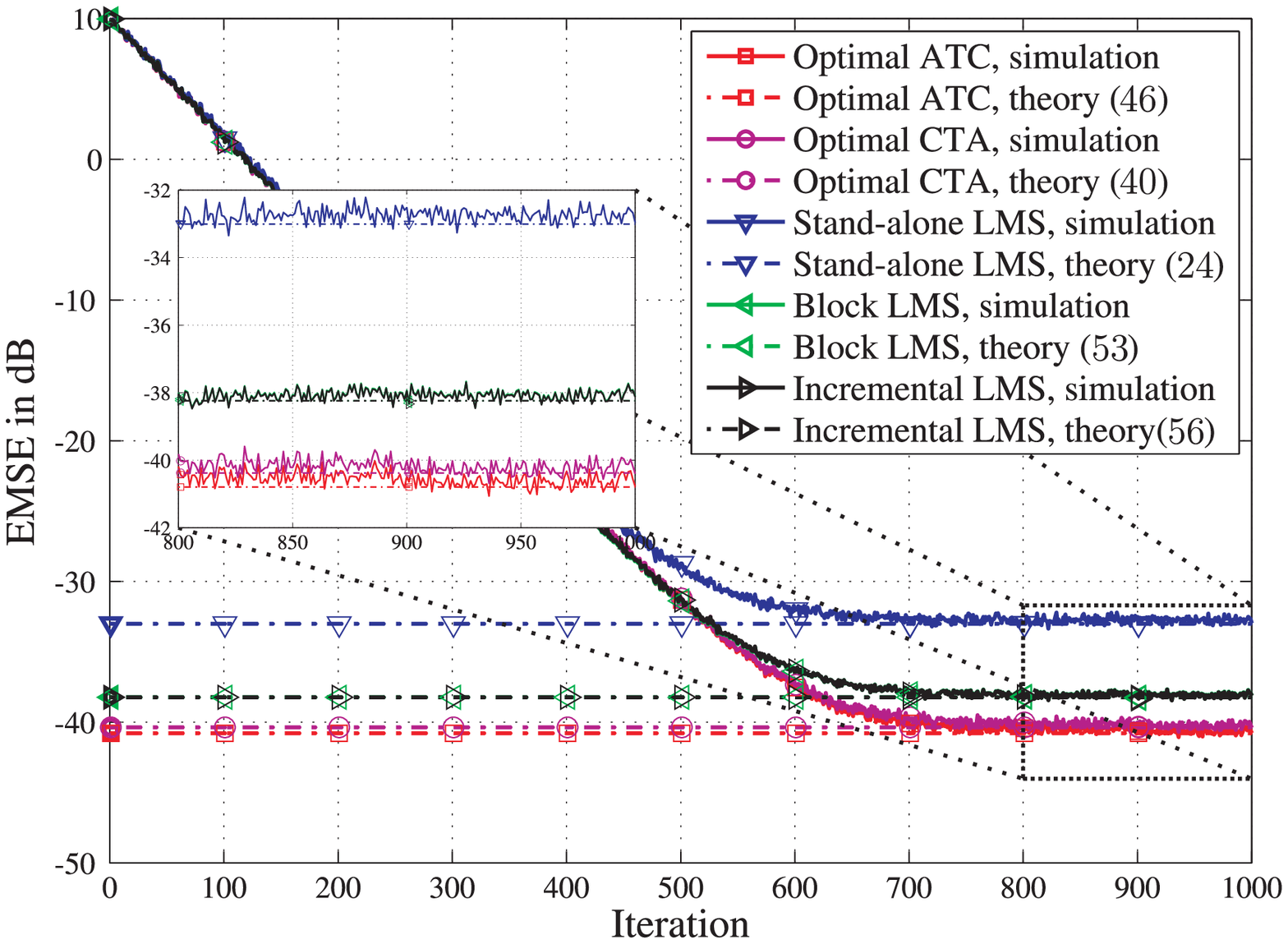}
\label{fig:node1}}
\hfil
\subfloat[Node 2.]
{\includegraphics[width=3.2in]{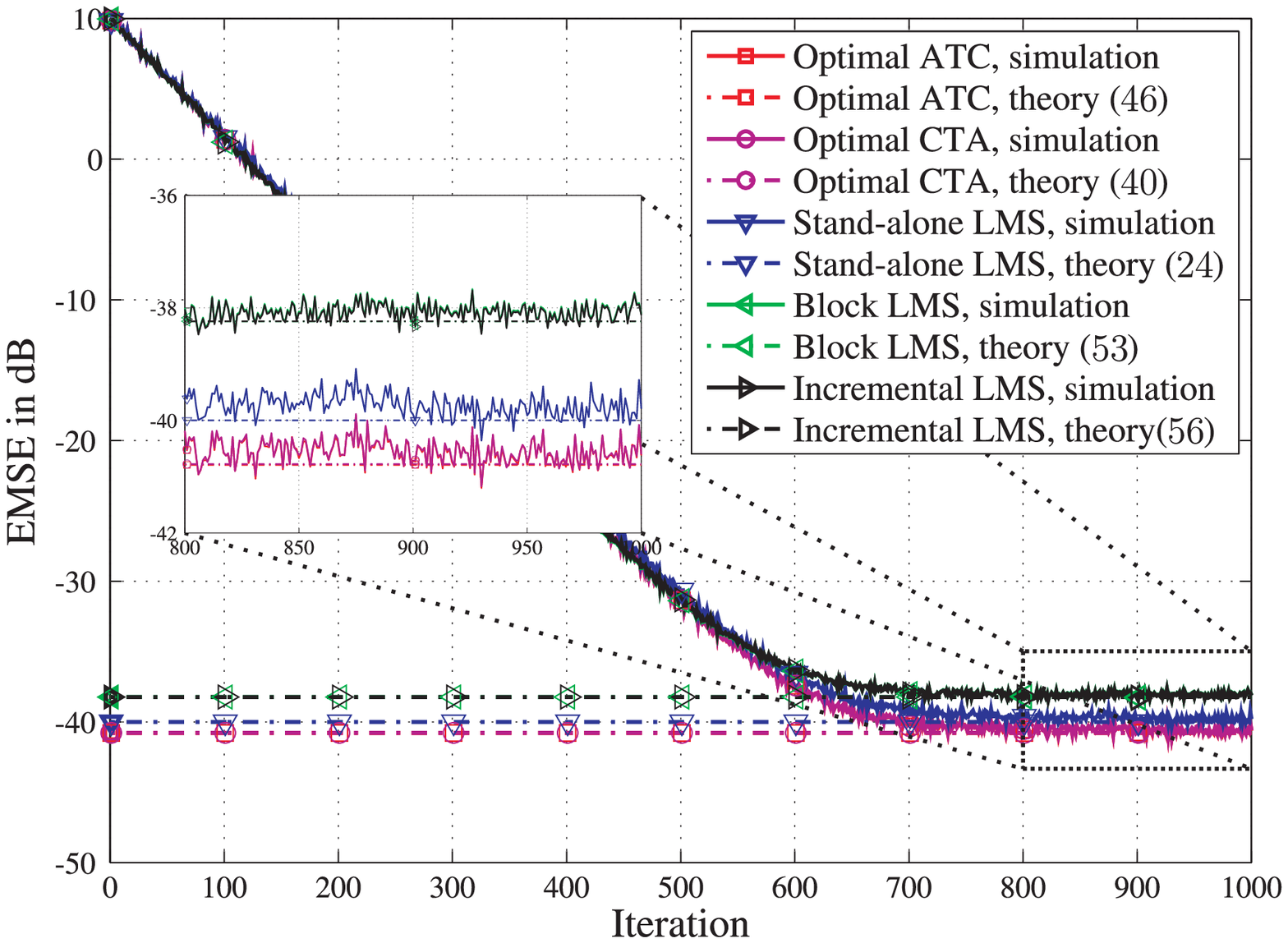}
\label{fig:node2}}}
\caption{Comparison of individual EMSE when $M=10$, $R_u=I_M$, $\sigma_{v,1}^2=0.01$, $\sigma_{v,2}^2=0.002$, and $\mu=0.01$.}
\label{fig:node}
\end{figure*}

We compare the network EMSE for various strategies in Fig. \ref{fig:network2nodes}. The length of $w^o$ is $M=10$ and its entries are randomly selected. The regression data $\{{\bs{u}}_{k,i}\}$ and noise signals $\{{\bs{v}}_k(i)\}$ are i.i.d. white Gaussian distributed with zero mean and $R_u=I_M$, $\sigma_{v,1}^2=0.01$, and $\sigma_{v,2}^2=0.002$. The results are averaged over 500 trials. From the simulation results, we can see that although centralized algorithms like \eqref{eqn:blockLMS} and \eqref{eqn:incrementalLMS} can offer a better estimate than the non-cooperative LMS algorithms \eqref{eqn:standaloneLMS}, they can be outperformed by the diffusion strategies \eqref{eqn:CTAnodek} and \eqref{eqn:ATCnodek} . When the combination coefficients of ATC or CTA algorithms are chosen according to the relative degree-variance rule \eqref{eqn:OptCondCTA}, these diffusion strategies can achieve lower EMSE by a significant margin. In addition, we compare the EMSE of each node in the network for various strategies in Figs. \ref{fig:node1}--\ref{fig:node2}.

\section{Performance of $N$-Node Adaptive Networks}
In the previous sections, we focused on two-node networks and were able to analytically characterize their performance, and to establish the superiority of the diffusion strategies over the centralized block or incremental LMS implementations. We now extend the results to $N$-node ad-hoc networks. First, we establish that for sufficiently small step-sizes and for any \emph{doubly-stochastic} combination matrix $A$, i.e., its rows and columns add up to one, the ATC diffusion strategy matches the performance of centralized block LMS. Second, we argue that by optimizing over the larger class of left-stochastic combination matrices, which include doubly-stochastic matrices as well, the performance of ATC can be improved relative to block LMS. Third, we provide a fully-distributed construction for the combination weights in order to minimize the network EMSE for ATC. We illustrate the results by focusing on ATC strategies but they apply to CTA strategies as well.

Thus, consider a connected network consisting of $N$-nodes. Each node $k$ collects measurement data that satisfy the linear regression model \eqref{eqn:linearmodel}. The noise variance at each node $k$ is $\sigma_{v,k}^2$. We continue to use Assumptions \ref{asm:all}--\ref{asm:uniform}. Each node $k$ runs the following ATC diffusion strategy \cite{Cattivelli10TSP}:
\begin{equation}
\left\{\begin{aligned}
{\bs{\psi}}_{k,i}&={\bs{w}}_{k,i-1}+\mu{\bs{u}}_{k,i}^*\left[{\bs{d}}_{k}(i)-{\bs{u}}_{k,i}{\bs{w}}_{k,i-1}\right]\\
{\bs{w}}_{k,i}&=\sum_{l\in{\mc{N}}_k}a_{lk}{\bs{\psi}}_{l,i}\\
\end{aligned}\right.
\end{equation}
where $a_{lk}$ denotes the positive weight that node $k$ assigns to data arriving from its neighbor $l$; these weights are collected into an $N\times N$ combination matrix $A$, and ${\mc{N}}_k$ consists of all neighbors of node $k$ including $k$ itself. The weights $\{a_{lk}\}$ satisfy the following properties:
\begin{align}
\label{eqn:topologyconstraint}
\sum_{l\in{\mc{N}}_k}a_{lk}=1,\quad a_{lk}>0\;\;{\textrm{if}}\;l\in{\mc{N}}_k,\;\;{\textrm{and}}\;\; a_{lk}=0\;\;{\textrm{if}}\;l\notin{\mc{N}}_k
\end{align}

\subsection{EMSE and MSD for ATC Diffusion LMS}
Observe that $A$ is a left-stochastic matrix (the entries on each of its columns add up to one). Let $A=TDT^{-1}$ denote the eigen-decomposition of $A$, where $T$ is a real and invertible matrix and $D$ is in the real Jordan canonical form \cite{Horn85,Laub05}. We assume that $A$ is a primitive/regular matrix, meaning that there exists an integer $m$ such that all entries of $A^m$ are strictly positive \cite{Horn85,BermanPF}. This condition essentially states that for any two nodes in the network, there is a path of length $m$ linking them. Since we assume a connected network and allow for loops because of \eqref{eqn:topologyconstraint}, it follows that $A$ satisfies the regularity condition \cite{Horn85,BermanPF,Sayed13Chapter}. Then, from the Perron-Frobenius theorem \cite{Horn85,BermanPF}, the largest eigenvalue in magnitude of $A$ is unique and is equal to one. Therefore, $D$ has the following form:
\begin{align}
\label{eqn:Ddef}
D=\begin{bmatrix}
1 \;&\; \\
 \;&\; J \\
\end{bmatrix}
\end{align}
where the $N-1\times N-1$ matrix $J$ consists of real stable Jordan blocks. From Appendix \ref{app:DerivationEMSE}, the network EMSE and MSD for ATC diffusion are given by
\begin{align}
\label{eqn:ndiffEMSEdefatc}
{\ol{\EMSE}}_{\atc}&\approx\frac{\mu^2}{N}\sum_{m=1}^{M}\lambda_m^2
{\mathrm{vec}}(D^{\T}T^{\T}R_vTD)^{\T}
(I_{N^2}-\xi_mD\otimes D)^{-1}{\mathrm{vec}}(T^{-1}T^{-\T})\\
\label{eqn:ndiffMSDdefatc}
{\ol{\MSD}}_{\atc}&\approx\frac{\mu^2}{N}\sum_{m=1}^{M}\lambda_m
{\mathrm{vec}}(D^{\T}T^{\T}R_vTD)^{\T}
(I_{N^2}-\xi_mD\otimes D)^{-1}{\mathrm{vec}}(T^{-1}T^{-\T})
\end{align}
where $\xi_m$ is given by \eqref{eqn:eigOmegaapprox} and
\begin{align}
\label{eqn:Rvdef}
R_v\defeq\diag\{\sigma_{v,1}^2,\dots,\sigma_{v,N}^2\}
\end{align}
From \eqref{eqn:eigOmegaapprox} and \eqref{eqn:Ddef}, we get
\begin{align}
\label{eqn:inversematrix}
&\mu(I_{N^2}-\xi_mD\otimes D)^{-1}=\begin{bmatrix}
(2\lambda_m)^{-1} & & & \\
& \mu(I-\xi_mJ)^{-1} & & \\
& & \mu(I-\xi_mJ)^{-1} & \\
& & & \mu(I-\xi_mJ\otimes J)^{-1} \\
\end{bmatrix}
\end{align}
where, to simplify the notation, we are omitting the subscripts of the identity matrices. Since $J$ is stable and $0<\xi_m<1$ under Assumption \ref{asm:smallstepsize}, we have
\begin{align}
\mu(I-\xi_mJ)^{-1}&=\mu(I-J+2\mu\lambda_mJ)^{-1}\approx\mu(I-J)^{-1}=O(\mu)\\
\mu(I-\xi_mJ\otimes J)^{-1}&=\mu(I-J\otimes J+2\mu\lambda_mJ\otimes J)^{-1}\approx\mu(I-J\otimes J)^{-1}=O(\mu)
\end{align}
Therefore, by Assumption \ref{asm:smallstepsize}, we can ignore all blocks on the diagonal of \eqref{eqn:inversematrix} with the exception of the left-most corner entry so that:
\begin{align}
\mu(I_{N^2}-\xi_mD\otimes D)^{-1}\approx(2\lambda_m)^{-1}E_{11}\otimes E_{11}
\end{align}
where $E_{11}$ now denotes the $N\times N$ matrix given by $E_{11}=\diag\{1,0,0,\dots,0\}$. Then,
\begin{align}
\label{eqn:rankone}
&\mu\,{\mathrm{vec}}(D^{\T}T^{\T}R_vTD)^{\T}(I_{N^2}-\xi_mD\otimes D)^{-1}{\mathrm{vec}}(T^{-1}T^{-\T})\nonumber\\
{}&\qquad\qquad\approx{\mathrm{vec}}(D^{\T}T^{\T}R_vTD)^{\T}[(2\lambda_m)^{-1}E_{11}\otimes E_{11}]{\mathrm{vec}}(T^{-1}T^{-\T})\nonumber\\
{}&\qquad\qquad=(2\lambda_m)^{-1}{\mathrm{vec}}(R_v)^{\T}(TE_{11}T^{-1}\otimes TE_{11}T^{-1}){\mathrm{vec}}(I_N)
\end{align}
where we used the fact that $DE_{11}=E_{11}$ because of \eqref{eqn:Ddef} and $\vecm(ABC)=(C^\T\otimes A)\vecm(B)$ for matrices $\{A,B,C\}$ of compatible dimensions. Now, note that $TE_{11}T^{-1}$ is a rank-one matrix determined by the outer product of the left- and right-eigenvectors of $A$ corresponding to the unique eigenvalue at one. Since $A$ is left-stochastic, this left-eigenvector can be selected as the all-one vector ${\ds{1}}$, i.e., $A^{\T}{\ds{1}}={\ds{1}}$. Let us denote the right-eigenvector by $y$ and normalize its element-sum to one, i.e., $Ay=y$ and $y^{\T}{\ds{1}}=1$. It follows from the Perron-Frobenius theorem \cite{Horn85,BermanPF} that all entries of $y$ are nonnegative and located within the range $[0,1]$. We then get $TE_{11}T^{-1}=y{\ds{1}}^{\T}$. Thus, from \eqref{eqn:rankone}, the network EMSE \eqref{eqn:ndiffEMSEdefatc} can be rewritten as
\begin{align}
{\ol{\EMSE}}_{\atc}&\approx\frac{\mu\Tr(R_u)}{2N}\vecm(R_v)^{\T}(y{\ds{1}}^{\T}\otimes y{\ds{1}}^{\T})\vecm(I_N)\nonumber\\
{}&=\frac{\mu\Tr(R_u)}{2N}\vecm(R_v)^{\T}\vecm(y{\ds{1}}^{\T}{\ds{1}}y^{\T})
\end{align}
That is,
\begin{equation}
\label{eqn:ndiffEMSEdefatcrankone}
\boxed{
{\ol{\EMSE}}_{\atc}\approx\frac{\mu\Tr(R_u)}{2}y^{\T}R_vy
}\end{equation}
Similarly, the network MSD \eqref{eqn:ndiffMSDdefatc} can be rewritten as
\begin{equation}
\label{eqn:ndiffMSDdefatcrankone}
\boxed{
{\ol{\MSD}}_{\atc}\approx\frac{\mu M}{2}y^{\T}R_vy
}\end{equation}

\subsection{EMSE and MSD for Block and Incremental LMS}
For $N$-nodes, the block LMS recursion \eqref{eqn:blockLMS} is replaced by
\begin{align}
\label{eqn:blockLMSnew}
{\bs{w}}_i={\bs{w}}_{i-1}+\mu'\sum_{k=1}^{N}{\bs{u}}_{k,i}^*\left[{\bs{d}}_{k}(i)-{\bs{u}}_{k,i}{\bs{w}}_{i-1}\right]
\end{align}
and the incremental LMS recursion \eqref{eqn:incrementalLMS} is replaced by
\begin{align}
\label{eqn:incrementalLMSnew}
\begin{cases}
\mbox{for every $i$:} \\
\quad{\mbox{initialize with}} \;\; {\bs{\psi}}_{0}={\bs{w}}_{i-1} \\
\quad{\mbox{for every $k=1,2,\dots,N$, repeat}}: \\
\quad\quad{\bs{\psi}}_k={\bs{\psi}}_{k-1}+\mu'{\bs{u}}_{k,i}^*\left[{\bs{d}}_{k}(i)-{\bs{u}}_{k,i}{\bs{\psi}}_{k-1}\right]\\
\quad\mbox{set} \;\; {\bs{w}}_{i}={\bs{\psi}}_N \\
\mbox{end}
\end{cases}
\end{align}
In order for block and incremental LMS to converge at the same rate as diffusion ATC, we must set their step-sizes to $\mu'=\mu/N$ (compare with \eqref{eqn:mu1andmu2}). Following an argument similar to the one presented in Appendix \ref{app:DerivationEMSEvec}, we can derive the EMSE and MSD for the block LMS strategy \eqref{eqn:blockLMSnew} as
\begin{equation}
\label{eqn:nEMSEveclms}
\boxed{
{\EMSE}_{\blk}\approx\frac{\mu\Tr(R_u)}{2}\frac{\Tr(R_v)}{N^2}
}\end{equation}
and
\begin{equation}
\label{eqn:nMSDveclms}
\boxed{
{\MSD}_{\blk}\approx\frac{\mu M}{2}\frac{\Tr(R_v)}{N^2}
}\end{equation}
respectively. A similar argument to \eqref{eqn:incremental2} (see also expression (84) in \cite{Lopes07TSP}) leads to the conclusion that the performance of incremental LMS \eqref{eqn:incrementalLMSnew} can be well approximated by that of block LMS for small step-sizes\footnote{Again, we remark that in general incremental LMS outperforms block LMS \cite{Cattivelli11TSPincremental}; however, their performance are similar when the step-size is sufficiently small \cite[App. A]{Lopes07TSP}.}. Therefore,
\begin{equation}
\label{eqn:nEMSEinclms}
\boxed{
{\EMSE}_{\inc}\approx\frac{\mu\Tr(R_u)}{2}\frac{\Tr(R_v)}{N^2}
}\end{equation}
and
\begin{equation}
\label{eqn:nMSDinclms}
\boxed{
{\MSD}_{\inc}\approx\frac{\mu M}{2}\frac{\Tr(R_v)}{N^2}
}\end{equation}
For this reason, we shall not distinguish between block LMS and incremental LMS in the sequel.

\subsection{Comparing Network EMSE}
Observe that the EMSE expression \eqref{eqn:ndiffEMSEdefatcrankone} for ATC diffusion LMS and \eqref{eqn:nEMSEveclms} for block and incremental LMS only differ by a scaling factor, namely, $y^{\T}R_vy$ versus $\Tr\left(R_v\right)/{N^2}$. Then, ATC diffusion would outperform block LMS and incremental LMS when
\begin{align}
y^{\T}R_vy<\frac{\Tr(R_v)}{N^2}
\end{align}
where $R_v$ is diagonal and given by \eqref{eqn:Rvdef}. We assume that the noise variance of at least one node in the network is different from the other noise variances to exclude the case in which the noise profile is uniform across the network (in which case $R_v$ would be a scaled multiple of the identity matrix). Thus, note that, if we select the combination matrix $A$ to be \emph{doubly-stochastic}, i.e., $A{\ds{1}}={\ds{1}}$ and $A^{\T}{\ds{1}}={\ds{1}}$, then it is straightforward to see that $y={\ds{1}}/N$ so that
\begin{align}
\label{eqn:doublychoice}
y^{\T}R_vy=\frac{\Tr(R_v)}{N^2}
\end{align}
This result means that, for sufficiently small step-sizes and for any doubly-stochastic matrix $A$, the EMSE performance of ATC diffusion and block LMS match each other. However, as indicated by \eqref{eqn:topologyconstraint}, the diffusion LMS strategy can employ a broader class of combination matrices, namely, left-stochastic matrices. If we optimize over the larger set of left-stochastic combination matrices and in view of \eqref{eqn:doublychoice}, we would expect
\begin{align}
{\EMSE}_{\blk}&\approx{\ol{\EMSE}}_{\atc}(A^{\textrm{doubly-stochastic}})
\geq{\ol{\EMSE}}_{\atc}(A^{\opt})
\end{align}
where $A^{\opt}$ is the optimal combination matrix that solves the following optimization problem:
\begin{equation}
\label{eqn:optproblem}
\begin{aligned}
&A^{\opt}\defeq\arg\min_{A\in{\mathbb{A}}}\;\frac{\mu\Tr(R_u)}{2}y^{\T}R_vy \\
{}&\quad \st\quad Ay=y,\;\;{\ds{1}}^{\T}y=1 \\
\end{aligned}
\end{equation}
where ${\mathbb{A}}$ denotes the set consisting of all $N\times N$ left-stochastic matrices whose entries $\{a_{lk}\}$ satisfy the conditions in \eqref{eqn:topologyconstraint}. We show next how to determine left-stochastic matrices that solve \eqref{eqn:optproblem}.

First note that the optimization problem \eqref{eqn:optproblem} is equivalent to the following non-convex problem:
\begin{equation}
\begin{aligned}
\label{eqn:optimalProb1}
\minimize_{A\in{\mathbb{A}},\;y\in{\mathbb{R}}_+} & \qquad\quad y^\T R_vy \\
\st & \quad Ay=y, \quad {\ds{1}}^\T y=1 \\
\end{aligned}
\end{equation}
where ${\mathbb{R}}_+$ denotes the $N\times1$ nonnegative vector space. We solve this problem in two steps. First we solve the  \emph{relaxed} problem:
\begin{equation}
\begin{aligned}
\label{eqn:optimalProb2}
\minimize_{y\in{\mathbb{R}}_+} & \quad y^\T R_vy \\
\st & \quad {\ds{1}}^\T y=1 \\
\end{aligned}
\end{equation}
Since $R_v$ is positive definite and diagonal, the closed-form solution for \eqref{eqn:optimalProb2} is given by
\begin{equation}
\label{eqn:optimalydef}
\boxed{
y^o\defeq\frac{R_v^{-1}{\ds{1}}}{{\ds{1}}^\T R_v^{-1}{\ds{1}}}
}\end{equation}
Next, if we can determine a primitive left-stochastic matrix $A$ whose right eigenvector associated to eigenvalue 1 coincides with $y^o$, then we would obtain a solution to \eqref{eqn:optimalProb1}. Indeed, note that any primitive left-stochastic matrix $A$ can be regarded as the probability transition matrix of an irreducible aperiodic Markov chain (based on the connected topology and condition \eqref{eqn:topologyconstraint} on the weights) \cite{Papoulis02,Meyn09}. In that case, a vector $y^o$ that satisfies $Ay^o=y^o$ would correspond to the stationary distribution vector for the Markov chain. Now given an arbitrary vector $y^o$, whose entries are positive and add up to one, it is known how to construct a left-stochastic matrix $A$ that would satisfy $Ay^o=y^o$. A procedure due to Hastings \cite{Hastings70Bio} was used in \cite{Boyd04SIAM} to construct such matrices. Applying the procedure to our vector $y^o$ given by \eqref{eqn:optimalydef}, we arrive at the following combination rule, which we shall refer to as the Hastings rule (we may add that there are many other choices for $A$ that would satisfy the same requirement $Ay^o=y^o$):
\begin{equation}
\label{eqn:MHrule}
{\mbox{\emph {Hastings rule:}}}\qquad\boxed{
a_{lk}=\begin{cases}
\displaystyle \frac{\sigma_{v,k}^2}{\max\{|\N_k|\sigma_{v,k}^2,|\N_l|\sigma_{v,l}^2\}}, & l\in\N_k\backslash\{k\} \\
\displaystyle 1-\sum_{l\in\N_k\backslash\{k\}}a_{lk}, & l=k \\
\end{cases}
}\end{equation}
where $|{\mc{N}}_k|$ denotes the cardinality of ${\mc{N}}_k$. It is worth noting that the Hastings rule is a \emph{fully-distributed} solution --- each node $k$ only needs to obtain the degree-variance product $(|\N_l|-1)\sigma_{v,l}^2$ from its neighbor $l$ to compute the corresponding combination weight $a_{lk}$. By using the Hastings rule \eqref{eqn:MHrule}, the vector $y^o$ in \eqref{eqn:optimalydef} is attained and the EMSE expression \eqref{eqn:ndiffEMSEdefatcrankone} is therefore minimized. The minimum value of \eqref{eqn:ndiffEMSEdefatcrankone} is then given by
\begin{align}
{\ol{\EMSE}}_{\atc}^{\opt}&\approx\frac{\mu\Tr(R_u)}{2}y^{o\T}R_vy^o
=\frac{\mu\Tr(R_u)}{2}\frac{1}{\Tr(R_v^{-1})}
\end{align}
Compared to the EMSE of block and incremental LMS \eqref{eqn:nEMSEveclms} and \eqref{eqn:nEMSEinclms}, we conclude that diffusion strategies using the Hastings rule \eqref{eqn:MHrule} achieve a lower EMSE level under Assumption \ref{asm:smallstepsize}. This is because, from the Cauchy-Schwarz inequality \cite{Sayed08}, we have
\begin{align}
N^2<\Tr(R_v)\Tr(R_v^{-1})\Longleftrightarrow\frac{1}{\Tr(R_v^{-1})}<\frac{\Tr(R_v)}{N^2}
\end{align}
when the entries on the diagonal of $R_v$ are not uniform (as we assumed at the beginning of this subsection).

\begin{figure*}[t]
\centerline{
\subfloat[Network topology and noise profile.]
{\includegraphics[height=2.5in]{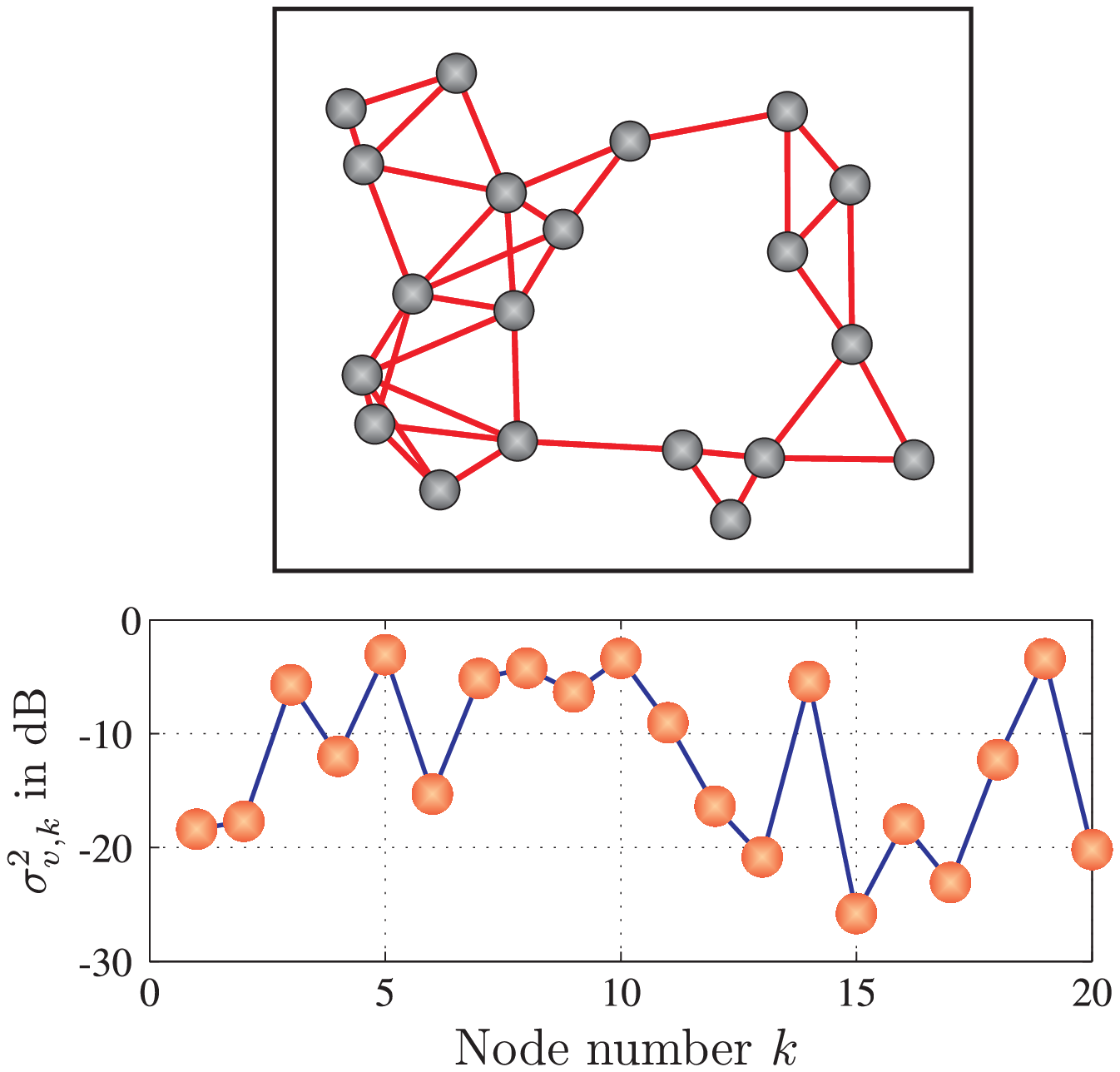}
\label{fig:profiles}}
\hfil
\subfloat[EMSE curves.]
{\includegraphics[height=2.5in]{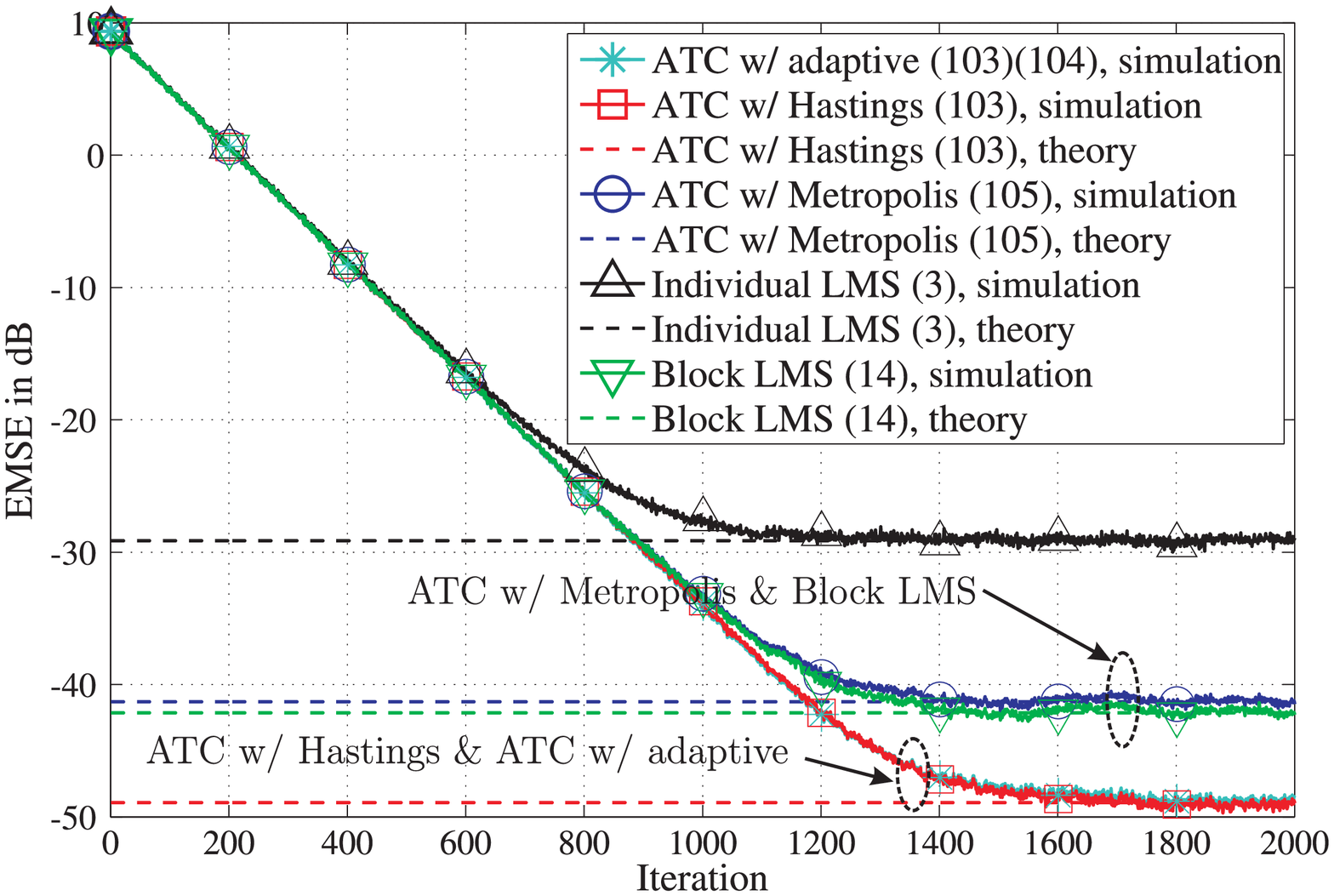}
\label{fig:EMSE}}}
\caption{Simulated EMSE curves and theoretical results for ATC diffusion versus block LMS for a network with $N=20$ nodes.}
\label{fig:sim20nodes}
\end{figure*}

In real applications, where the noise variances are unavailable, each node can estimate its own noise variance recursively by using the following iteration:
\begin{align}
\label{eqn:estimatenoisevariance}
\wh{\bm{\sigma}}_{v,k}^2(i)=(1-\nu_k)\wh{\bm{\sigma}}_{v,k}^2(i-1)+\nu_k|\bm{d}_k(i)-\bm{u}_{k,i}\bm{w}_{k,i-1}|^2
\end{align}
\noindent\emph{Remark}: In the two-node case, we determined the combination weights \eqref{eqn:OptCondCTA} by seeking coefficients that essentially minimize the EMSE expressions \eqref{eqn:EMSEctanetwork} and \eqref{eqn:EMSEatcnetwork}. The argument in Appendix \ref{app:MinimizeEMSEcta} expressed the EMSE as the sum of two factors: a dominant factor that depends on $\mu$ and a less dominant factor that depends on higher powers of $\mu$. In the $N$-node network case, we instead used the small step-size approximation to arrive at expressions \eqref{eqn:ndiffEMSEdefatcrankone} and \eqref{eqn:ndiffMSDdefatcrankone}, which correspond only to the dominant terms in the EMSE and MSD expressions and depend on $\mu$. We can regard \eqref{eqn:ndiffEMSEdefatcrankone} and \eqref{eqn:ndiffMSDdefatcrankone} as first-order approximations for the performance of the network for sufficiently small step-sizes.

\subsection{Simulation Results}
We simulate ATC diffusion LMS versus block LMS over a connected network with $N=20$ nodes. The unknown vector $w^o$ of length $M=3$ is randomly generated. We adopt $R_{u}=I_M$, $\mu=0.005$ for ATC diffusion LMS, and $\mu'=\mu/N=0.00025$ for block LMS. The network topology and the profile of noise variances $\{\sigma_{v,k}^2\}$ are plotted in Fig. \ref{fig:profiles}. For ATC algorithms, we simulate three different combination rules: the first one is the (left-stochastic) adaptive Hastings rule \eqref{eqn:MHrule} using \eqref{eqn:estimatenoisevariance} and without the knowledge of noise variances, the second one is the Hastings rule \eqref{eqn:MHrule} with the knowledge of noise variance, and the third one is the (doubly-stochastic) Metropolis rule \cite{Metropolis53JCP,Cattivelli10TSP} (which is a simplified version of the Hastings rule):
\begin{equation}
\label{eqn:Metropolisrule}
\mbox{\emph {Metropolis rule:}}\qquad\boxed{a_{lk}=\begin{cases}
\displaystyle \frac{1}{\max\{|{\mc{N}}_k|,|{\mc{N}}_l|\}}, & l\in\N_k\backslash\{k\} \\
\displaystyle 1-\sum_{l\in\N_k\backslash\{k\}}a_{lk}, & l=k \\
\end{cases}}
\end{equation}
We adopted $\nu_k=0.1$ and $\mu=0.0054$ for the adaptive Hastings rule \eqref{eqn:MHrule}--\eqref{eqn:estimatenoisevariance} to match the convergence rate of the other algorithms. We also consider the non-cooperative LMS case for comparison purposes. The EMSE learning curves are obtained by averaging over 50 experiments and are plotted in Fig. \ref{fig:EMSE}. It can be seen that ATC diffusion LMS with Metropolis weights exhibits almost the same convergence behavior as block LMS in transient phase and attains a steady-state value that is less than 1 dB worse than block LMS. In comparison, ATC diffusion LMS using adaptive Hastings weights (where the noise variances are estimated through \eqref{eqn:estimatenoisevariance}) has almost the same learning curve as ATC using Hastings weights with the knowledge of the noise variances; both of them are able to attain about 7 dB gain over block LMS at steady-state.

\section{Conclusion}
In this work we derived the EMSE levels for different strategies over LMS adaptive networks and compared their performance. The results establish that diffusion LMS strategies can deliver lower EMSE than centralized solutions employing traditional block or incremental LMS strategies. We first studied the case of networks involving two cooperating nodes, where closed-form expressions for the EMSE and MSD can be derived. Subsequently, we extended the conclusion to generic $N$-node networks and established again that, for sufficiently small step-sizes, diffusion strategies can outperform centralized block LMS strategies by optimizing over left-stochastic combination matrices. It is worth noting that although the optimized combination rules rely on knowledge of the noise statistics, it is possible to employ adaptive strategies like \eqref{eqn:estimatenoisevariance} to adjust these coefficients on the fly without requiring explicit knowledge of the noise profile --- in this way, the Hastings rule \eqref{eqn:MHrule} can be implemented in a manner similar to the adaptive relative variance rule \cite{Tu11CAMSAP,Zhao12TSP,Sayed13Chapter}. Clearly, the traditional block and incremental implementations \eqref{eqn:blockLMSnew} and \eqref{eqn:incrementalLMSnew} can be modified to incorporate information about the noise profile as well. In that case, it can be argued that diffusion strategies are still able to match the EMSE performance of these modified centralized algorithms.

\appendices

\section{EMSE Expression for General Diffusion LMS with $N$-Nodes}
\label{app:DerivationEMSE}
Under Assumptions \ref{asm:all}--\ref{asm:uniform}, the EMSE expression for node $k$ of the general diffusion strategy \eqref{eqn:idealDiffusionPriorDiff}--\eqref{eqn:idealDiffusionPostDiff} is given by Eq. (39) from reference \cite{Cattivelli10TSP} (see also \cite{Sayed13Chapter}):
\begin{align}
\label{eqn:EMSEkdef}
  {\EMSE}_k\approx[\vecm({\mc{Y}}^\T)]^\T(I_{N^2M^2} - {\mc{F}})^{-1}\vecm(E_{kk}\otimes R_{u})
\end{align}
where
\begin{align}
\label{eqn:Ydef}
{\mc{Y}}&=\mu^2(Q^{\T}R_vQ)\otimes R_u \\
\label{eqn:Fdef}
{\mc{F}}&={\mc{B}}^{\T}\otimes{\mc{B}}^* \\
\label{eqn:Bdef}
{\mc{B}}&=(Q^{\T}P^{\T})\otimes(I_M-\mu R_u)
\end{align}
and $E_{kk}=\diag\{0,\dots,0,1,0,\dots,0\}$ is an $N\times N$ all-zero matrix except for the $k$th entry on the diagonal, which is equal to one. Since for ATC algorithms, $P=I_N$ and $Q=A$, and for CTA algorithms, $P=A$ and $Q=I_N$, we know that $PQ=A$ for both cases. Therefore, we get
\begin{align}
{\mc{B}}=A^{\T}\otimes(I_M-\mu R_u)
\end{align}
We can reduce \eqref{eqn:EMSEkdef} into the form \eqref{eqn:diffEMSEdefnew}, which is more suitable for our purposes, by introducing the eigen-decompositions of $R_u$ and $A$. Thus, let $R_u={U}\Lambda{U}^*$ denote the eigen-decomposition of $R_u$, where ${U}$ is unitary and $\Lambda$ is diagonal with positive entries. Let also $A=TDT^{-1}$ denote the eigen-decomposition of the real matrix $A$, where $T$ is real and invertible and $D$ is in the \emph{real} Jordan canonical form \cite{Horn85,Laub05}. Then, the eigen-decomposition of ${\mc{B}}$ is given by
\begin{align}
{\mc{B}}&=(TDT^{-1})^{\T}\otimes[{U}(I_M-\mu\Lambda){U}^*]\nonumber\\
{}&=(T^{-\T}\otimes{U})[D^{\T}\otimes(I_M-\mu\Lambda)](T^{\T}\otimes{U}^*)
\end{align}
and the eigen-decomposition of ${\mc{F}}$ is then given by
\begin{align}
\label{eqn:Fdefnew}
{\mc{F}}&=\{(T^{-\T}\otimes{U})[D^{\T}\otimes(I_M-\mu\Lambda)](T^{\T}\otimes{U}^*)\}^{\T}
\otimes\{(T^{-\T}\otimes{U})[D^{\T}\otimes(I_M-\mu\Lambda)](T^{\T}\otimes{U}^*)\}^*\nonumber\\
{}&={\mc{X}}\{[D^{\T}\otimes(I_M-\mu\Lambda)]^{\T}\otimes[D^{\T}\otimes(I_M-\mu\Lambda)]^*\}{\mc{X}}^{-1}\nonumber\\
{}&={\mc{X}}\{[D\otimes(I_M-\mu\Lambda)]\otimes[D\otimes(I_M-\mu\Lambda)]\}{\mc{X}}^{-1}\nonumber\\
{}&={\mc{X}}({\mc{G}}\otimes{\mc{G}}){\mc{X}}^{-1}
\end{align}
where we used the facts that $\{D,\Lambda\}$ are real and $\Lambda$ is diagonal, and introduced the matrices:
\begin{align}
\label{eqn:Xdef}
{\mc{X}}&\defeq(T^{\T}\otimes{U}^*)^{\T}\otimes(T^{\T}\otimes{U}^*)^*\\
\label{eqn:Gdef}
{\mc{G}}&\defeq D\otimes(I_M-\mu\Lambda)
\end{align}
Then, from \eqref{eqn:Ydef}--\eqref{eqn:Xdef}, we get
\begin{align}
\label{eqn:XYeqn}
{\mc{X}}^\T\vecm({\mc{Y}}^\T)&=[(T^{\T}\otimes{U}^*)\otimes(T^{*}\otimes{U}^\T)]\cdot\mu^2\vecm(Q^{\T}R_vQ\otimes R_u^\T)\nonumber\\
&=\mu^2\vecm\left[(T^{*}\otimes{U}^\T)(Q^{\T}R_vQ\otimes R_u^\T)(T\otimes{U}^{*\T})\right]\nonumber\\
&=\mu^2\vecm(T^{\T}Q^{\T}R_vQT\otimes\Lambda)
\end{align}
where we used the fact that $T$ is real. Likewise, we get
\begin{align}
\label{eqn:XinvEeqn}
{\mc{X}}^{-1}\vecm(E_{kk}\otimes R_{u})&=[(T^{-\T}\otimes{U})^{\T}\otimes(T^{-\T}\otimes{U})^*]\cdot\vecm(E_{kk}\otimes R_{u})\nonumber\\
&=\vecm(T^{-1}E_{kk}T^{-\T}\otimes\Lambda)
\end{align}
Then, from \eqref{eqn:Fdefnew}--\eqref{eqn:XinvEeqn}, the EMSE expression in \eqref{eqn:EMSEkdef} can be rewritten as
\begin{align}
\label{eqn:EMSEkdef1}
{\textrm{EMSE}}_k&\approx[\vecm({\mc{Y}}^\T)]^\T{\mc{X}}(I_{N^2M^2}-{\mc{G}}\otimes{\mc{G}})^{-1}
{\mc{X}}^{-1}\vecm(E_{kk}\otimes R_{u})\nonumber\\
{}&=\mu^2[\vecm(T^{\T}Q^{\T}R_vQT\otimes\Lambda)]^{\T}(I_{N^2M^2}-{\mc{G}}\otimes{\mc{G}})^{-1}
\vecm(T^{-1}E_{kk}T^{-\T}\otimes\Lambda)
\end{align}
Using the fact that ${\mc{G}}$ in \eqref{eqn:Gdef} is stable under Assumption \ref{asm:smallstepsize}, we can further obtain
\begin{align}
\label{eqn:EMSEkdef2}
{\textrm{EMSE}}_k&\approx\mu^2[\vecm(T^{\T}Q^{\T}R_vQT\otimes\Lambda)]^{\T}
\left(\sum_{j=0}^{\infty}{\mc{G}}^j\otimes{\mc{G}}^j\right)
\vecm(T^{-1}E_{kk}T^{-\T}\otimes\Lambda)\nonumber\\
{}&=\mu^2[\vecm(T^{\T}Q^{\T}R_vQT\otimes\Lambda)]^{\T}
\sum_{j=0}^{\infty}\vecm\left[{\mc{G}}^j(T^{-1}E_{kk}T^{-\T}\otimes\Lambda){\mc{G}}^{\T j}\right]\nonumber\\
{}&=\mu^2\sum_{j=0}^{\infty}\Tr\left[(T^{\T}Q^{\T}R_vQT\otimes\Lambda){\mc{G}}^j
(T^{-1}E_{kk}T^{-\T}\otimes\Lambda){\mc{G}}^{\T j}\right]
\end{align}
where we used the identities $\vecm(ABC)=(C^\T\otimes A)\vecm(B)$ and $\Tr(AB)=[\vecm(A^\T)]^\T\vecm(B)$ for matrices $\{A,B,C\}$ of compatible dimensions. From \eqref{eqn:Xdef}, we get
\begin{align}
\label{eqn:tracesimplify}
&\Tr\left[(T^{\T}Q^{\T}R_vQT\otimes\Lambda){\mc{G}}^j(T^{-1}E_{kk}T^{-\T}\otimes\Lambda){\mc{G}}^{\T j}\right]\nonumber\\
{}&\quad=\Tr\left\{(T^{\T}Q^{\T}R_vQT\otimes\Lambda)[D^j\otimes(I_M-\mu\Lambda)^j]
(T^{-1}E_{kk}T^{-\T}\otimes\Lambda)[D^{\T j}\otimes(I_M-\mu\Lambda)^{j}]\right\}\nonumber\\
{}&\quad=\Tr\left[T^{\T}Q^{\T}R_vQTD^jT^{-1}E_{kk}T^{-\T}D^{\T j}
\otimes\,\Lambda(I_M-\mu\Lambda)^j\Lambda(I_M-\mu\Lambda)^{j}\right]\nonumber\\
{}&\quad=\Tr\left[\Lambda(I_M-\mu\Lambda)^j\Lambda(I_M-\mu\Lambda)^{j}
\otimes\,T^{\T}Q^{\T}R_vQTD^jT^{-1}E_{kk}T^{-\T}D^{\T j}\right]\nonumber\\
{}&\quad=\sum_{m=1}^{M}\lambda_m^2(1-\mu\lambda_m)^{2j}
\Tr(T^{\T}Q^{\T}R_vQTD^jT^{-1}E_{kk}T^{-\T}D^{\T j})\nonumber\\
{}&\quad=\sum_{m=1}^{M}\lambda_m^2(1-\mu\lambda_m)^{2j}[\vecm(T^{\T}Q^{\T}R_vQT)]^\T
(D^{j}\otimes D^j)\vecm(T^{-1}E_{kk}T^{-\T})
\end{align}
where we used the identity $\Tr(A\otimes B)=\Tr(B\otimes A)$ for square matrices $\{A,B\}$ and the fact that $\Lambda(I_M-\mu\Lambda)^j\Lambda(I_M-\mu\Lambda)^{j}$ is diagonal. Substituting \eqref{eqn:tracesimplify} back into \eqref{eqn:EMSEkdef2} leads to
\begin{align}
\label{eqn:EMSEkdef1new}
{\textrm{EMSE}}_k&\approx\mu^2\sum_{m=1}^{M}\lambda_m^2[\vecm(T^{\T}Q^{\T}R_vQT)]^\T
\left[\sum_{j=0}^{\infty}(1 - \mu\lambda_m)^{2j}D^{j}\otimes D^j\right]
\vecm(T^{-1}E_{kk}T^{-\T})\nonumber\\
{}&=\mu^2\sum_{m=1}^{M}\lambda_m^2[\vecm(T^{\T}Q^{\T}R_vQT)]^{\T}
[I_{N^2} - (1 - \mu\lambda_m)^2D\otimes D]^{-1}{\mathrm{vec}}(T^{-1}E_{kk}T^{-\T})\nonumber\\
{}&\approx\mu^2\sum_{m=1}^{M}\lambda_m^2[\vecm(T^{\T}Q^{\T}R_vQT)]^{\T}
[I_{N^2} - (1 - 2\mu\lambda_m)D\otimes D]^{-1}{\mathrm{vec}}(T^{-1}E_{kk}T^{-\T})
\end{align}
where $(1-\mu\lambda_m)^2\approx1-2\mu\lambda_m$ due to Assumption \ref{asm:smallstepsize}.

\section{Minimizing the Network Performance for Diffusion LMS}
\label{app:MinimizeEMSEcta}
To minimize the network EMSE for CTA given by \eqref{eqn:EMSEctanetwork}, we introduce two auxiliary variables $\eta$ and $\theta$ such that $\alpha+\beta=1+\eta$ and $1-\beta=\theta(1-\alpha)$, where $-1\le\eta<1$ and $\theta>0$. The network EMSE \eqref{eqn:EMSEctanetwork} can be rewritten as
\begin{align}
\label{eqn:app3EMSEapprox}
{\ol{\EMSE}}_{\cta}&\approx\mu^2\sigma_{v,1}^2\sum_{m=1}^{M}\frac{\lambda_m^2}{(1+\theta)^2}
\left[\frac{\theta^2+\gamma}{1-\xi_m}
+\frac{(1-\theta)(\theta-\gamma)}{1-\xi_m\eta}
+\frac{1+\gamma}{2}\frac{1+\theta^2}{1-\xi_m\eta^2}\right]
\end{align}
where $\gamma\defeq\sigma_{v,2}^2/\sigma_{v,1}^2<1$ and $0<\xi_m<1$ is given by \eqref{eqn:ximdef} under Assumption \ref{asm:smallstepsize}. Minimizing expression \eqref{eqn:app3EMSEapprox} in closed-form over both variables $\{\theta,\eta\}$ is generally non-trivial. We exploit the fact that the step-size is sufficiently small to help locate the values of $\theta$ and $\eta$ that approximately minimize the value of \eqref{eqn:app3EMSEapprox}. For this purpose, we first substitute \eqref{eqn:ximdef} into \eqref{eqn:app3EMSEapprox} and use Assumption \ref{asm:smallstepsize} to note that
\begin{align}
\label{eqn:app3EMSEapproxnew}
{\ol{\EMSE}}_{\cta}&\approx\frac{\mu\Tr(R_u)\sigma_{v,1}^2}{2}\frac{\theta^2+\gamma}{(1+\theta)^2}+O(\mu^2)
\end{align}
Expression \eqref{eqn:app3EMSEapproxnew} writes the EMSE as the sum of two factors: the first factor is linear in the step-size and depends only on $\theta$, and the second factor depends on higher-order powers of the step-size. For sufficiently small step-sizes, the first factor is dominant and we can ignore the second factor. Doing so allows us to estimate the value of $\theta$ that minimizes \eqref{eqn:app3EMSEapprox}. Observe that the first factor on RHS of \eqref{eqn:app3EMSEapproxnew} is minimized at $\theta^o=\gamma$ because
\begin{align}
\gamma^2+\theta^2\ge2\theta\gamma 
\;\;\Longrightarrow\;\;  \theta^2\gamma+\gamma+\gamma^2+\theta^2\ge\theta^2\gamma+\gamma+2\theta\gamma 
\;\;\Longrightarrow\;\;  \frac{\theta^2+\gamma}{(1+\theta)^2}\ge\frac{\gamma}{1+\gamma}
\end{align}
We now substitute $\theta^o=\gamma$ back into the original expression \eqref{eqn:app3EMSEapprox} for the network EMSE to find that:
\begin{align}
\label{eqn:app3EMSEapproxnew2}
{\ol\EMSE}_{\cta}&\approx\mu^2\sigma_{v,1}^2\sum_{m=1}^{M}\frac{\lambda_m^2}{1+\gamma}
\left[\frac{\gamma}{1-\xi_m}+\frac{1+\gamma^2}{2(1-\xi_m\eta^2)}\right]
\end{align}
We use this form to minimize the higher-order terms of $\mu$ over the variable $\eta$. It is obvious that expression \eqref{eqn:app3EMSEapproxnew2} is minimized at $\eta^o=0$. The value of EMSE under $\theta^o=\gamma$ and $\eta^o=0$ is then given by
\begin{align}
\label{eqn:app3EMSEapprox3}
{\ol\EMSE}_{\cta}(\theta^o=\gamma,\eta^o=0)
\approx\sigma_{v,1}^2\left[\frac{\gamma}{1+\gamma}\frac{\mu\Tr(R_u)}{2}+\frac{1+\gamma^2}{2(1+\gamma)}\sum_{m=1}^{M}\mu^2\lambda_m^2\right]
\end{align}
Similarly, we can employ the same approximate argument to find that the solution $(\theta^o,\eta^o)$ essentially minimizes the network MSD under Assumption \ref{asm:smallstepsize}; the corresponding value of the MSD is
\begin{align}
\label{eqn:app3MSDapprox3}
{\ol\MSD}_{\cta}(\theta^o=\gamma,\eta^o=0)
\approx\sigma_{v,1}^2\left[\frac{\gamma}{1+\gamma}\frac{\mu M}{2}
+\frac{1+\gamma^2}{1+\gamma}\frac{\mu^2\Tr(R_u)}{2}\right]
\end{align}
The solution $\{\theta^o=\gamma,\eta^o=0\}$ translates into \eqref{eqn:OptCondCTA}, where $0<\gamma<1$.

In a similar manner, in order to minimize the network EMSE of ATC given by \eqref{eqn:EMSEatcnetwork}, we introduce two auxiliary variables $\eta$ and $\theta$ such that $\alpha+\beta=1+\eta$ and $1-\beta=\theta(1-\alpha)$, where $-1\le\eta<1$ and $\theta>0$. Then, from \eqref{eqn:EMSEatcnetwork} we have
\begin{align}
{\ol\EMSE}_{\atc}\approx\mu^2\sigma_{v,1}^2\sum_{m=1}^{M}\frac{\lambda_m^2}{\xi_m}\left[\frac{1}{(1+\theta)^2}
\left(\frac{\theta^2+\gamma}{1-\xi_m}+\frac{(1-\theta)(\theta-\gamma)}{1-\xi_m\eta}+\frac{1+\gamma}{2}\frac{1+\theta^2}{1-\xi_m\eta^2}\right)
-\frac{1+\gamma}{2}\right]
\end{align}
for which we can again motivate the selection $\{\theta^o=\gamma,\eta^o=0\}$. The value of the network EMSE at $\theta^o=\gamma$ and $\eta^o=0$ is then given by
\begin{align}
{\ol{\EMSE}}_{\atc}(\theta^o=\gamma,\eta^o=0)&\approx\sigma_{v,1}^2\frac{\gamma}{1+\gamma}\frac{\mu\Tr(R_u)}{2}
\end{align}

\section{Derivation of EMSE for Block LMS Networks}
\label{app:DerivationEMSEvec}
We start from \eqref{eqn:blockLMS}. To simplify the notation, we rewrite \eqref{eqn:linearmodel} and \eqref{eqn:blockLMS}  as
\begin{align}
{\bs{d}}_i&={\bs{U}}_iw^o+{\bs{v}}_i\\
{\bs{w}}_{i}&={\bs{w}}_{i-1}+\mu{\bs{U}}_i^*({\bs{d}}_i-{\bs{U}}_i{\bs{w}}_{i-1})
\end{align}
where
\begin{align}
{\bs{U}}_i&\defeq{\col}\{{\bs{u}}_{1,i},{\bs{u}}_{2,i}\}\\
{\bs{d}}_i&\defeq{\col}\{{\bs{d}}_{1}(i),{\bs{d}}_{2}(i)\}\\
{\bs{v}}_i&\defeq{\col}\{{\bs{v}}_{1}(i),{\bs{v}}_{2}(i)\}
\end{align}
The error recursion is then given by
\begin{align}
\label{eqn:app6esterrrecurdef}
{\widetilde{\bs{w}}}_{i}=(I_M-\mu{\bs{U}}_i^*{\bs{U}}_i){\widetilde{\bs{w}}}_{i-1}-\mu{\bs{U}}_i^*{\bs{v}}_i
\end{align}
Let $\Sigma$ be an arbitrary $M\times M$ positive semi-definite matrix that we are free to choose. Using \eqref{eqn:app6esterrrecurdef}, we can evaluate the weighted square quantity $\|{\widetilde{\bs{w}}}_{i}\|_{\Sigma}^{2}\equiv{\wt{\bm{w}}}_i^*\Sigma{\wt{\bm{w}}}_i$. Doing so and taking expectations under Assumption \ref{asm:all}, we arrive at the following weighted variance relation \cite{Sayed08,Cattivelli11TSPincremental}:
\begin{align}
\label{eqn:app6energy3}
\E\|{\widetilde{\bs{w}}}_{i}\|_{\Sigma}^{2}&=\E\|{\widetilde{\bs{w}}}_{i-1}\|_{\Sigma'}^{2}
+\mu^2\E\|{\bs{U}}_i^*{\bs{v}}_i\|_{\Sigma}^{2}
\end{align}
where
\begin{align}
\label{eqn:app6Sigmaprime}
\Sigma'&\triangleq\E(I_M-\mu{\bs{U}}_i^*{\bs{U}}_i)\Sigma(I_M-\mu{\bs{U}}_i^*{\bs{U}}_i)\nonumber\\
{}&\approx\Sigma-2\mu R_u\Sigma-2\mu\Sigma R_u
\end{align}
where, in view of Assumption \ref{asm:smallstepsize}, we are dropping higher-order terms in $\mu$. Let again $R_u=U\Lambda U^*$ denote the eigen-decomposition of $R_u$. We then introduce the transformed quantities:
\begin{alignat}{2}
{\ol{\bs{w}}}_i&\defeq{U}^*{\bs{w}}_i, &\qquad
{\ol{\bs{U}}}_i&\defeq{\bs{U}}_i{U} \\
{\ol{\Sigma}}&\defeq{U}^*\Sigma{U}, &\qquad
{\ol{\Sigma}}'&\defeq{U}^*\Sigma'{U}
\end{alignat}
Relation \eqref{eqn:app6energy3} is accordingly transformed into
\begin{align}
\label{eqn:app6transformedenergy3}
\E\|{\ol{\bs{w}}}_{i}\|_{\ol{\Sigma}}^{2}&=\E\|{\ol{\bs{w}}}_{i-1}\|_{{\ol\Sigma}'}^{2}
+\mu^2\E\|{\ol{\bs{U}}}_i^*{\bs{v}}_i\|_{\ol{\Sigma}}^{2}
\end{align}
where
\begin{align}
{\ol{\Sigma}}'&\approx{\ol{\Sigma}}-2\mu\Lambda{\ol{\Sigma}}-2\mu{\ol{\Sigma}}\Lambda
\end{align}
Since we are free to choose $\Sigma$, or equivalently, ${\ol{\Sigma}}$, let ${\ol{\Sigma}}$ be diagonal and nonnegative. Then, it can be verified that ${\ol{\Sigma}}'$ is also diagonal and nonnegative under Assumptions \ref{asm:all}--\ref{asm:uniform} so that
\begin{align}
{\ol{\Sigma}}'\approx(I_M-4\mu\Lambda){\ol{\Sigma}}
\end{align}
Under Assumption \ref{asm:all}, the second term on the right-hand side of \eqref{eqn:app6transformedenergy3} evaluates to
\begin{align}
\mu^2\E\|{\ol{\bs{U}}}_i^*{\bs{v}}_i\|_{\ol{\Sigma}}^{2}&=\mu^2\Tr[R_v(\E\,{\bs{U}}_i\Sigma{\bs{U}}_i^*)]
\end{align}
where
\begin{align}
\E\,{\bs{U}}_i\Sigma{\bs{U}}_i^*&=\E\begin{bmatrix}
{\bs{u}}_{1,i}\Sigma{\bs{u}}_{1,i}^* & {\bs{u}}_{1,i}\Sigma{\bs{u}}_{2,i}^* \\
{\bs{u}}_{2,i}\Sigma{\bs{u}}_{1,i}^* & {\bs{u}}_{2,i}\Sigma{\bs{u}}_{2,i}^* \\
\end{bmatrix}=\Tr({\overline{\Sigma}}\Lambda)I_2
\end{align}
Therefore, we get
\begin{align}
\mu^2{\mathbb{E}}\|{\overline{\bs{U}}}_i^*{\bs{v}}_i\|_{\overline{\Sigma}}^{2}=\mu^2\Tr(R_v)\Tr({\overline{\Sigma}}\Lambda)
\end{align}
When the filter is mean-square stable, taking the limit as $i\rightarrow\infty$ of both sides of \eqref{eqn:app6transformedenergy3} and selecting $\ol{\Sigma}=I_M/4\mu$, we get
\begin{align}
\label{eqn:app6EMSEdef}
{\EMSE}_{\blk}\approx\frac{\mu\Tr(R_u)}{2}\frac{\sigma_{v,1}^2+\sigma_{v,2}^2}{2}
\end{align}
Likewise, by selecting $\ol{\Sigma}=\Lambda^{-1}/4\mu$ and taking the limit of both sides of \eqref{eqn:app6transformedenergy3} as $i\rightarrow\infty$, we arrive at
\begin{align}
\label{eqn:app6MSDdef}
{\MSD}_{\blk}\approx\frac{\mu M}{2}\frac{\sigma_{v,1}^2+\sigma_{v,2}^2}{2}
\end{align}

\section*{Acknowledgment}
The authors would like to acknowledge useful feedback from Ph.D. student Z. Towfic on Sec. VI-C.

\end{document}